\documentclass{article}
\usepackage[margin=1.25in]{geometry}
\usepackage{graphicx}
\graphicspath{{FigBSSSvelodrome_arXiv/}}
\usepackage{float}
\usepackage{amsmath}
\usepackage{amssymb}
\usepackage{mathtools}
\usepackage{parskip}
\usepackage[hidelinks]{hyperref}
\usepackage{graphicx}
\usepackage[bf]{caption}
\usepackage{subcaption}

\usepackage{color}
\usepackage{natbib}
\usepackage{amssymb}
\usepackage{ulem}

\usepackage{booktabs}
\emergencystretch=\maxdimen
\usepackage{tikz}
\usetikzlibrary{calc,angles,quotes} 
\usetikzlibrary{arrows.meta}
\usepackage{caption}
\usepackage{subcaption}
\begin{document}
\title{
	Modelling of a cyclist's power for time trials on a velodrome
}
\author{
	Len Bos%
	\footnote{
		Universit\`a di Verona, Italy, \texttt{leonardpeter.bos@univr.it}
	}\,,
	Michael A. Slawinski%
	\footnote{
		Memorial University of Newfoundland, Canada, \texttt{mslawins@mac.com}
	}\,,
	Rapha\"el A. Slawinski%
	\footnote{
		Mount Royal University, Canada, \texttt{rslawinski@mtroyal.ca}
	}\,,
	Theodore Stanoev%
	\footnote{
		Memorial University of Newfoundland, Canada, \texttt{theodore.stanoev@gmail.com}
	}
}
\date{}
\maketitle
\begin{abstract}
\noindent We formulate a phenomenological model to study the power applied by a cyclist on a velodrome\,---\,for individual timetrials\,---\,taking into account the straights, circular arcs, connecting transition curves and banking.
The dissipative forces we consider are air resistance, rolling resistance, lateral friction and drivetrain resistance.
Also, power can be used to increase the kinetic and gravitational potential energy.
Herein, to model a steady ride\,---\,as expected for individual  timetrials\,---\,we assume a constant centre-of-mass speed, while allowing the cadence and power to vary during a lap.
Hence, the kinetic energy is constant and the only mechanical energy whose change we need to consider is the increase of gravitational potential energy due to raising the centre of mass upon exiting each curve.
The effect of dissipative forces is examined at each point of the lap; the effect of conservative forces is examined as an average.
The latter is a small\,---\,albeit not negligible\,---\,part of the total power, and its inclusion within a model is a novelty presented herein.
It increases the model's empirical adequacy.

Following derivations and justifications of expressions that constitute this mathematical model, we present a numerical example.
We show that the cadence and power vary slightly during a steady ride.
In other words, a constant centre-of-mass speed entails nearly constant cadence and power, as expected for a steady ride and as supported by measurements.

Also, we examine changes in the required power due to changes of various quantities, such as air density at a velodrome, laptime and several others, as well as the model sensitivity to input errors.
Furthermore, we examine the effects on the required power of slight and gradual changes in speed, which are pertinent to individual time trials.

Such examinations are of immediate use in strategizing the performance for individual pursuits and the Hour Record, as well as in evaluating agreements between models and measurements since both are subject to inaccuracies and errors.
Notably, scenarios generated with our model were taken into consideration prior to two successful Hour Records in 2022: Daniel Bigham and Filippo Ganna.

This article contains six appendices.
Therein, we show aspects of derivations and discuss the effects of chosen approximations.
\end{abstract}
\section{Introduction}
In this article, we formulate a mathematical model to examine the power expended by a cyclist on a velodrome.
The model includes several simplifying assumptions that permit the derivation of closed-form expressions which lend themselves to inverse problems, while nevertheless capturing the mechanical behaviour in an empirically adequate manner.
The assumptions pertain to both the bicycle-cyclist system and the design of the track.

For the former, we assume the cyclist to be already in a launched effort, which renders the model adequate for longer events, such as a 4000-metre pursuit or, in particular, the Hour Record.
Limiting our attention to a steady ride, we view its motion as tantamount to a constant centre-of-mass speed.
To consider the effect of leaning, we distinguish between the centre-of-mass and wheel speeds to decompose the forces that act on various parts of the bicycle-cyclist system throughout a given lap.
We determine the lean angles by assuming the system is in instantaneous rotational equilibrium about the line of contact with the tires on the ground, which is tantamount to net zero torque, about the axis parallel to the instantaneous velocity through the centre of mass.

For the latter, geometrically, we consider a velodrome with its straights, circular arcs and connecting transition curves, whose inclusion\,---\,as indicated by \citet{FitzgeraldEtAl2021}\,---\,has been commonly neglected in previous studies.
While this inclusion presents a mathematical challenge, it increases the empirical adequacy of the model.
However, we consider that the geometric features of the black line, which is an offset curve at a fixed distance from the inner edge of the track, are adequately represented by a curve confined to the horizontal plane.
In addition, we assume an idealized wheel path such that its trajectory coincides with the black line.
We focus our attention on phenomenological quantities that affect the bicycle-cyclist system, with an emphasis on the required power.

Various power models and velodrome geometries have been presented in other studies.
Almost a quarter-of-a-century ago, \citet{MartinEtAl1998} formulate and examine a road-cycling model.
Proceeding with velodrome models, \citet{UnderwoodJermy2010,UnderwoodJermy2014} formulate and examine a model for individual pursuits that includes leaning on the bends, but only uses straights and bends without transition curves. 
\citet{LukesEtAl2012} model the power of a cyclist on a velodrome, including the tire scrubbing effects.
\citet{FittonSymons2018} present a model that includes slip and steering angles and variable black-line altitude, and use velodrome models based on theodolite measurements.
\citet{Stanoev2023} models a velodrome as a ruled surface and provides technical considerations of its design.

Recently, \citet{FitzgeraldEtAl2021} quantify the effects of transition curves on a power model consistent with \citet{LukesEtAl2012} and \citet{MartinEtAl1998}.
They consider two types of transition curves, namely, a twice-differentiable linear increase in curvature, which is the Euler spiral, and a sinusoid, which exhibits a continuous derivative of curvature.
The effects of these curves are estimated using least squares optimization of black-line measurements obtained with theodolite measurements.

We begin this article by expressing mathematically the geometry of the black line and the inclination of the track.
We provide a detailed derivation of the Euler spiral, which already appeared as an earlier version of this arXiv and was referred to by~\citet{FitzgeraldEtAl2021} as the most detailed to date. 
Also, we derive a Bloss-transition curve, whose curvature is thrice-differentiable. 
Our expressions are accurate representations of the common geometry of modern $250$\,-metre velodromes, whose design details were provided by Mehdi Kordi ({\it pers.\!~comm.}, 2020).
We proceed to formulate an expression for power used against dissipative forces and power used to increase mechanical energy.
In particular, while earlier work considers changes in gravitational potential energy as a function of vertical location on a banked velodrome (e.g., \citet{BenhamEtAl2020}) or of the lean angle and the vertical inclination of trajectory (e.g., \citet{FittonSymons2018}), we consider\,---\,in accordance with the assumption of instantaneous rotational equilibrium\,---\,increases in gravitational potential energy due solely to straightening of the cyclist upon exiting the curves, with wheels remaining along the black line.
We show that a significant amount of power is expended in that process.
To present a numerical example and for a comparison with \citet{FitzgeraldEtAl2021}, we examine the case of a constant centre-of-mass speed, which we consider the best mathematical analogy for the steadiness of timetrial efforts on a velodrome.
As the last part of this article, we consider the applicability of our formulation to slight changes in speed.

For completeness, we include six appendices.
Therein, we present derivations of expressions for the effects of air resistance on power requirements, for the lean angle as a function of speed and radius of curvature, for change of gravitational potential energy with the lean angle, comparison of potential and rotational wheel energies as a function of the lean angle, effects of gradual speed increase on power requirements, and a brief discussion of empirical adequacy of model assumptions.
\section{Track}
\label{sec:Formulation}
\subsection{Black-line parameterization}
\label{sub:Track}
\begin{figure}[h]
\centering
\includegraphics[scale=0.5]{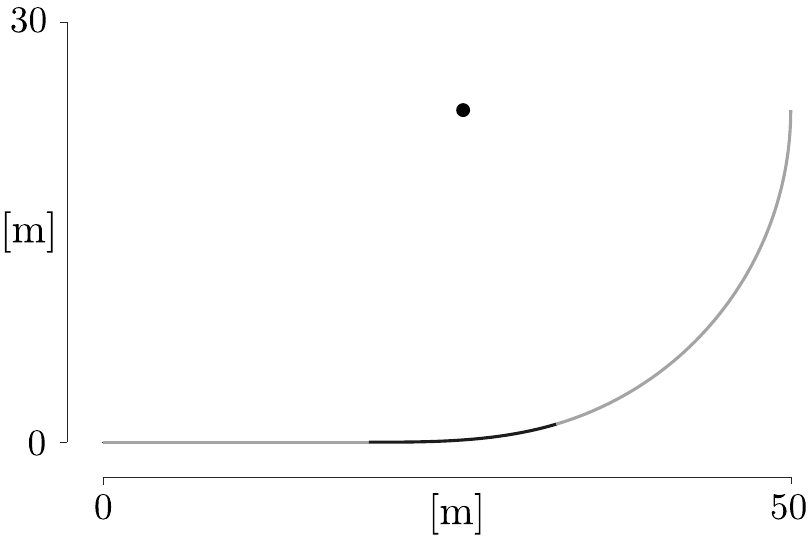}
\caption{\small  A quarter of the black line for a $250$\,-metre track}
\label{fig:FigTrack}
\end{figure}

To model the power required of a cyclist\,---\,riding along the black line\,---\,to achieve a given result, we define this line by three parameters.
\begin{itemize}
\item[$\circ$] $L_s$\,: the half-length of the straight
\item[$\circ$] $L_t$\,: the length of the transition curve between the straight and the circular arc
\item[$\circ$] $L_a$\,: the length of the remainder of the quarter circular arc
\end{itemize}
The length of the track is $S=4(L_s+L_t+L_a)$\,.
In Figure~\ref{fig:FigTrack}, we show a quarter of a black line for $L_s=19\,\rm m$\,, $L_t=13.5\,\rm m$ and $L_a=30\,\rm m$\,, which results in $S=250\,\rm m$\,, as required.
Therein, the transition curve is shown in black and connects the straight and the  circular arc, shown in gray.
This curve has continuous derivative up to order two; in other words, it is a $C^2$ curve.

To express, in Cartesian coordinates, the curve shown in Figure~\ref{fig:FigTrack}, we consider the following.
\begin{itemize}
\item[$\circ$] The straight,
\begin{equation*}
y_1=0\,,\qquad0\leqslant x\leqslant a\,,	
\end{equation*}
where $a:=L_s$\,.
\item[$\circ$] The transition, where\,---\,following a standard design practice\,---\,we take it to be an Euler spiral, which can be parameterized by Fresnel integrals,
\begin{equation*}
x_2(\varsigma)=a+\sqrt{\dfrac{2}{A}}\int\limits_0^{\varsigma\sqrt{\!\frac{A}{2}}}\!\!\!\!\cos\!\left(x^2\right)\,{\rm d}x
\qquad{\rm and}\qquad
y_2(\varsigma)=\sqrt{\dfrac{2}{A}}\int\limits_0^{\varsigma\sqrt{\!\frac{A}{2}}}\!\!\!\!\sin\!\left(x^2\right)\,{\rm d}x\,,
\end{equation*}
with $A>0$ to be determined; herein, $\varsigma$ is the curve parameter.
Since the arclength differential,~${\rm d}s$\,, is such that
\begin{equation*}
{\rm d}s=\sqrt{x_2'(\varsigma)^2+y_2'(\varsigma)^2}\,{\rm d}\varsigma
=\sqrt{\cos^2\left(\dfrac{A\varsigma^2}{2}\right)+\sin^2\left(\dfrac{A\varsigma^2}{2}\right)}\,{\rm d}\varsigma
={\rm d}\varsigma\,,
\end{equation*}
we write the transition curve as
\begin{equation*}
(x_2(s),y_2(s)), \quad 0\leqslant s\leqslant b:=L_t\,.
\end{equation*}
\item[$\circ$] The circular arc, whose centre is $(c_1,c_2)$ and whose radius is $R$\,, with $c_1$\,, $c_2$ and $R$  to be determined.
Since its arclength is specified to be $c:=L_a,$ we may parameterize the quarter circle by
\begin{equation}
\label{eq:x3}
x_3(\theta)=c_1+R\cos(\theta)
\end{equation}
and
\begin{equation}
\label{eq:y3}
y_3(\theta)=c_2+R\sin(\theta)\,,
\end{equation}
where $-\theta_0\leqslant\theta\leqslant 0$\,, for $\theta_0:=c/R$\,.
The centre of the circle is shown as a black dot in Figure~\ref{fig:FigTrack}.
\end{itemize}
We wish to connect these three segments so that the resulting track be continuous along with its first and second derivatives.

To do so, let us consider the connection between the straight and the transition curve.
Herein, $x_2(0)=a$ and $y_2(0)=0$\,, so the curve connects continuously to the end of the straight at $(a,0)$\,.
Also, at~$(a,0)$\,,
\begin{equation*}
\dfrac{{\rm d}y}{{\rm d}x}=\dfrac{y_2'(0)}{x_2'(0)}=\dfrac{0}{1}=0\,,
\end{equation*}
which matches the derivative of the straight line.
Furthermore, the second derivatives match, since
\begin{equation*}
\dfrac{{\rm d}^2y}{{\rm d}x^2}=\dfrac{y''_2(0)x_2'(0)-y'_2(0)x_2''(0)}{(x_2'(0))^2}=0\,,
\end{equation*}
which follows, for any $A>0$\,, from
\begin{equation}
\label{eq:FirstDer}
x_2'(\varsigma)=\cos\left(\dfrac{A\,\varsigma^2}{2}\right)\,, \quad y_2'(\varsigma)=\sin\left(\dfrac{A\,\varsigma^2}{2}\right)
\end{equation}
and
\begin{equation*}
x_2''(\varsigma)=-A\,\varsigma\sin\left(\dfrac{A\,\varsigma^2}{2}\right)\,, \quad y_2''(\varsigma)=A\,\varsigma\cos\left(\dfrac{A\,\varsigma^2}{2}\right)\,.
\end{equation*}
Let us consider also the connection between the transition curve and the arc of the circle. 
In order that these connect continuously,
\begin{equation*}
\big(x_2(b),y_2(b)\big)=\big(x_3(-\theta_0),y_3(-\theta_0)\big)\,,
\end{equation*}
we require
\begin{equation}
\label{eq:Cont1}
x_2(b)=c_1+R\cos(\theta_0)\,\,\iff\,\,c_1=x_2(b)-R\cos\!\left(\dfrac{c}{R}\right)
\end{equation}
and
\begin{equation}
\label{eq:Cont2}
y_2(b)=c_2-R\sin(\theta_0)\,\,\iff\,\, c_2=y_2(b)+R\sin\!\left(\dfrac{c}{R}\right)\,.
\end{equation}
For the tangents to connect continuously, we invoke expression~(\ref{eq:FirstDer}) to write
\begin{equation*}
(x_2'(b),y_2'(b))=\left(\cos\left(\dfrac{A\,b^2}{2}\right),\,\sin\left(\dfrac{A\,b^2}{2}\right)\right)\,.
\end{equation*}
Following expressions~(\ref{eq:x3}) and (\ref{eq:y3}), we obtain
\begin{equation*}
\big(x_3'(-\theta_0),y_3'(-\theta_0)\big)=\big(R\sin(\theta_0),R\cos(\theta_0)\big)\,,
\end{equation*}
respectively.
Matching the unit tangent vectors, we obtain
\begin{equation}
\label{eq:tangents}
\cos\left(\dfrac{A\,b^2}{2}\right)=\sin\!\left(\dfrac{c}{R}\right)\,,\quad \sin\left(\dfrac{A\,b^2}{2}\right)=\cos\!\left(\dfrac{c}{R}\right)\,.
\end{equation}
For the second derivative, it is equivalent\,---\,and easier\,---\,to match the curvature.
For the Euler spiral,
\begin{equation*}
\kappa_2(s)=\dfrac{x_2'(s)y_2''(s)-y_2'(s)x_2''(s)}
{\Big(\big(x_2'(s)\big)^2+\big(y_2'(s)\big)^2\Big)^{\frac{3}{2}}}
=A\,s\cos^2\left(\dfrac{A\,s^2}{2}\right)+A\,s\sin^2\left(\dfrac{A\,s^2}{2}\right)
=A\,s\,,
\end{equation*}
which is indeed the defining characteristic of an Euler spiral: the curvature grows linearly in the arclength.
Hence, to match the curvature of the circle at the connection, we require
\begin{equation*}
A\,b=\dfrac{1}{R} \,\,\iff\,\,A=\dfrac{1}{b\,R}\,.
\end{equation*}
Substituting this value of $A$ in equations~(\ref{eq:tangents}), we obtain
\begin{align*}
\cos\!\left(\dfrac{b}{2R}\right)&=\sin\!\left(\dfrac{c}{R}\right)\,,\quad \sin\!\left(\dfrac{b}{2R}\right)=\cos\!\left(\dfrac{c}{R}\right)\\
&\iff\dfrac{b}{2R}=\dfrac{\pi}{2}-\dfrac{c}{R}\\
&\iff R=\dfrac{b+2c}{\pi}.
\end{align*}
It follows that
\begin{equation*}
A=\dfrac{1}{b\,R}=\dfrac{\pi}{b\,(b+2c)}\,;
\end{equation*}
hence, the continuity condition stated in expressions~(\ref{eq:Cont1}) and (\ref{eq:Cont2}) determines the centre of the circle,~$(c_1,c_2)$\,.

\begin{figure}[h]
\centering
\includegraphics[scale=0.5]{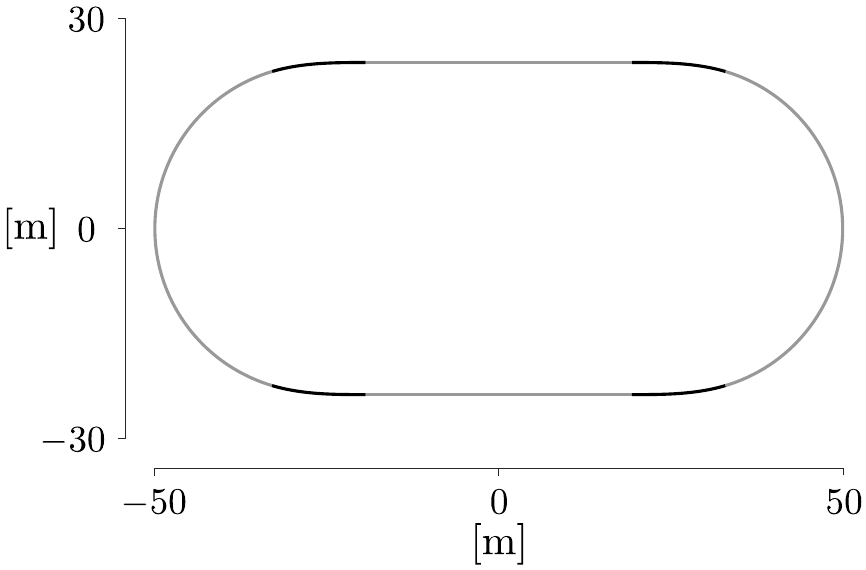}
\caption{\small Black line of $250$\,-metre track}
\label{fig:FigComplete}
\end{figure}

\begin{figure}[h]
\centering
\includegraphics[scale=0.5]{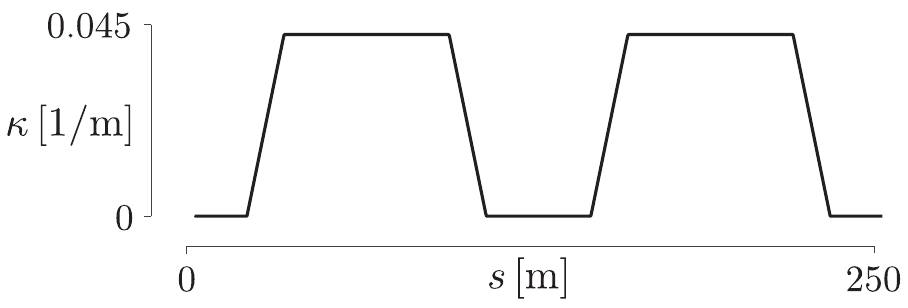}

\caption{\small Curvature of the black line,~$\kappa$\,, as a function of distance,~$s$\,, with a linear transition between the zero curvature of the straight and the $1/R$ curvature of the circular arc}
\label{fig:FigTrackCurvature}
\end{figure}

For the case shown in Figure~\ref{fig:FigTrack}, the numerical values are~$A=3.1661\times10^{-3}$\,m${}^{-2}$, $R=23.3958$\,m\,, $c_1=25.7313$\,m and $c_2=23.7194$\,m\,.
The complete track\,---\,with its centre at the origin\,,~$(0,0)$\,---\,is shown in Figure~\ref{fig:FigComplete}.
The corresponding track curvature is shown in Figure~\ref{fig:FigTrackCurvature}.
Note that  the curvature transitions linearly from the constant value of straight,~$\kappa=0$\,, to the constant value of the circular arc,~$\kappa=1/R$\,.
\subsection{$\boldsymbol{C^3}$ transition curve}
\label{sub:BiTraj}
As stated by~\cite{FitzgeraldEtAl2021},
\begin{quote}
\label{FitzQuote}
[t]here can be a trade-off between good spatial fit (low continuity for matching a built velodrome) and a good physical fit (high continuity for matching a cyclist's path).
If this model is used to calculate the motion of cyclists on a velodrome surface then a spatial fit can be prioritised.
However, if it is used to directly model the path of a cyclist then higher continuity is required.
\end{quote}
Should it be desired that the trajectory have continuous derivatives up to order $r\geqslant 0$\,---\,which makes it in class $C^r$\,, so that its curvature is of class $C^{r-2}$\,---\,it is possible to achieve this by a transition curve $y=p(x)$\,, where $p(x)$ is a polynomial of degree $2r-1$ and the solution of a $4\times 4$ system of nonlinear equations, assuming a solution exists.

In case $r=2$\,, this would correspond to a transition by a polynomial of degree $2r-1=3$\,, often referred to as a cubic spiral.
In case $r=3$\,, this would correspond to a transition by a polynomial of degree $2r-1=5$\,, often referred to as a Bloss transition.

The motivation for a $C^3$ transition curve, derived herein, is to increase the smoothness of the black line.
Since we assume this line to coincide with the wheel trajectory, the increased continuity increases the smoothness of power computed\,---\,in Section~\ref{sub:NumEx}, below\,---\,at discretized points along the black line.
A similar approach can be used to obtain curves of even higher continuity, should it be warranted by computations.

As in Section~\ref{sub:Track}, we let $a=L_s$ and we use a Cartesian formulation, where
\begin{itemize}
\item[$\circ$] $y=0,$ $0\leqslant x\leqslant a$\,: the halfstraight
\item[$\circ$] $y=p(x),$ $a\leqslant x\leqslant X$\,: the transition curve between the straight and circular arc
\item[$\circ$] $y=c_2-\sqrt{R^2-(x-c_1)^2}$\,: the end circle, with centre at $(c_1,c_2)$ and radius $R.$
\end{itemize}
Herein, $x=X$ is the point at which the polynomial transition connects to the circle, $(x-c_1)^2+(y-c_2)^2=R^2$\,.

$X$\,, $c_1$\,, $c_2$ and $R$ are {\it a priori} unknowns that need to be determined.
These four values determine completely the transition ,  $y=p(x),$ $a\leqslant x\leqslant X$\,.

Given $X$\,, $c_1$\,, $c_2$ and $R$\,, we can compute $p(x)$ as the Lagrange-Hermite interpolant of the data,
\[p^{(j)}(a)=0,\quad 0\leqslant j\leqslant r,\,\,\hbox{and}\,\, p^{(j)}(X)=y^{(j)}(X),\quad 0\leqslant j\leqslant r\,,\]
where $y(x)=c_2-\sqrt{R^2-(x-c_1)^2}$ is the circle.
Indeed, this is given in Newton form as
\[p(x)=\sum_{j=0}^{2r+1} p[x_0,\cdots,x_j]\prod_{k=0}^{j-1}(x-x_k)\,,\]
with 
\[x_0=x_1=\cdots=x_r=a\,\,\hbox{and},\,\, x_{r+1}=\cdots=x_{2r+1}=X\]
and, for any values $t_1,t_2,\cdots,t_m$\,, the divided difference\,,
\[p[t_1,t_2,\cdots,t_m]\,,\]
may be computed from the recurrence
\[p[t_1,\cdots,t_m]=\dfrac{p[t_2,\cdots.t_m]-p[t_1,\cdots,t_{m-1}]}{t_m-t_1}\,,\]
with
\[p\!\!\!\!\!\underbrace{[a,\cdots,a]}_{{\rm multiplicity}\,\,j+1}:=0\,,\,\, 0\leqslant j\leqslant r\]
and
\[p\!\!\!\underbrace{[X,\cdots,X]}_{{\rm multiplicity}\,\,j+1}:=\dfrac{y^{(j)}(X)}{j!}\,,\,\, 0\leqslant j\leqslant r\,,\]
where again $y(x)=c_2-\sqrt{R^2-(x-c_1)^2}$ is the circle.

Note, however, that this defines $p(x)$ as a polynomial of degree $2r+1$ exceeding our claimed degree $2r-1$ by 2. 
Hence, the first two of our equations are that $p(x)$ is actually of degree $2r-1$\,, thus, its two leading coefficients are zero.
More specifically, from the Newton form,
\begin{equation}\label{eq1}
p[\,\underbrace{a,a,\cdots,a}_{r+1\,\,{\rm copies}},\underbrace{X,X,\cdots,X}_{r+1\,\,{\rm copies}}\,]=0
\end{equation}
and
\begin{equation}\label{eq2}
p[\,\underbrace{a,a,\cdots,a}_{r+1\,\,{\rm copies}},\underbrace{X,X,\cdots,X}_{r\,\,{\rm copies}}\,]=0\,.
\end{equation}
The third equation is that the length of the transition is specified to be $L_t$\,; thus,
\begin{equation}\label{eq3}
\int\limits_a^X \sqrt{1+\left(p'(x)\right)^2}\,{\rm d}x=L_t\,.
\end{equation}
Note that this integral has to be estimated numerically.

Finally, the fourth equation is that the remainder of the circular arc\,---\,from $x=X$ to the apex $x=c_1+R$\,---\,has length $L_a$\,.
This is easily accomplished.
The two endpoints of the circular arc are
$(X,y(X))$ and $(c_1+R,c_2)$, giving a circular segment of angle $\theta$\,, say, between the two vectors connecting these points to the centre, namely, $\langle X-c_1,y(X)-c_2\rangle$ and $\langle R,0\rangle$\,.
Then,
\[\cos(\theta)=\dfrac{\langle X-c_1,y(X)-c_2\rangle\,\cdot\,\langle R,0\rangle} {\|\langle X-c_1,y(X)-c_2\rangle\|_2 \|\langle R,0\rangle\|_2}=\dfrac{X-c_1}{R}.\]
Hence the fourth equation is
\begin{equation}\label{eq4}
R\,\cos^{-1}\left(\dfrac{X-c_1}{R}\right)=L_a\,.
\end{equation}
Equations \eqref{eq1}, \eqref{eq2}, \eqref{eq3} and \eqref{eq4} form a system of four nonlinear equations in four unknowns, $c_1$\,, $c_2$\,, $R$ and $X$\,, determining a $C^r$ transition given by $p(x)$\,, a polynomial of degree $2r-1$\,, satisfying the design parameters.
\begin{figure}[h]
\centering
\includegraphics[scale=0.5]{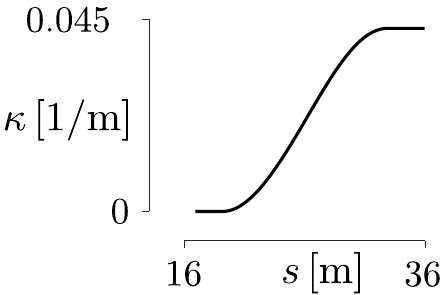}
\caption{\small Transition curvature,~$\kappa$\,, as a function of distance,~$s$\,, which\,---\,in contrast to Figure~\ref{fig:FigTrackCurvature}\,---\,exhibits a smooth ($C^1$) transition between the zero curvature of the straight and the $1/R$ curvature of the circular arc}
\label{fig:FigBloss}
\end{figure}

For the velodrome under consideration, and a Bloss $C^3$ transition, with curvature $C^1$\,, we obtain $c_1=25.7565\,\rm m$\,, $c_2=23.5753\,\rm m$\,, $R = 23.3863\,\rm m$ and $X =32.3988\,\rm m$\,, which, as expected, are similar\,---\,albeit not identical\,---\,to values obtained in Section~\ref{sub:Track}.
The transition curvature is shown in Figure~\ref{fig:FigBloss}.

Returning to the quote on page~\pageref{FitzQuote}, the requirement of differentiability of the trajectory traced by bicycle wheels, which is expressed in spatial coordinates, is a consequence of dynamic behaviour, which is expressed in temporal coordinates, namely, the requirement of smoothness of the derivative of acceleration, which is commonly referred to as a jolt.
In other words, a natural motion of the bicycle-cyclist system, which is laterally unconstrained, ensures that the third temporal derivative of position is smooth.
\subsection{Track-inclination angle}
\label{sub:TrackInc}
\begin{figure}[h]
\centering
\includegraphics[scale=0.5]{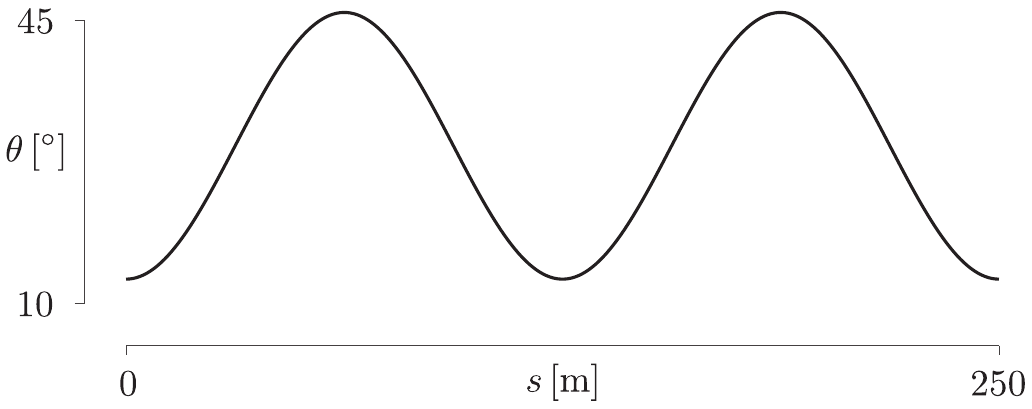}
\caption{\small Track inclination,~$\theta$\,, as a function of the black-line distance,~$s$}
\label{fig:FigAngle}
\end{figure}
To model the track-inclination angle, we choose a trigonometric formula in terms of arclength, which is a good analogy of an actual $250$\,-metre velodrome.
Herein, the minimum inclination of $13^\circ$ corresponds to the midpoint of the straight, and the maximum of $46^\circ$ to the apex of the circular arc; notably, such values are exhibited by the Tissot Velodrome in Grenchen, Switzerland.

For a track of length $S$\,,
\begin{equation}
\label{eq:theta}
\theta(s)=29.5-16.5\cos\!\left(\dfrac{4\pi}{S}s\right)\,;
\end{equation}
$s=0$ refers to the midpoint of the lower straight, in Figure~\ref{fig:FigComplete}, and the track is oriented in the counterclockwise direction.
Figure \ref{fig:FigAngle} shows this inclination for $S=250\,\rm m$\,.

It is not uncommon for tracks to be slightly asymmetric with respect to the inclination angle.
In such a case, they exhibit a rotational symmetry by $\pi$\,, but not a reflection symmetry about the vertical or horizontal axis.
This asymmetry can be modelled by including $s_0$ in the argument of the cosine in expression~(\ref{eq:theta}),
\begin{equation}
\label{eq:s0}
\theta(s)=29.5-16.5\cos\!\left(\dfrac{4\pi}{S}(s-s_0)\right)\,;
\end{equation}
$s_0\approx 5$ provides a good model for several existing velodromes.
Referring to discussions about the London Velodrome of the 2012 Olympics, \citet{Solarczyk} writes that
\begin{quote}
the slope of the track going into and out of the turns is not the same.
This is simply because you always cycle the same way around the track, and you go shallower into the turn and steeper out of it.
\end{quote}
The last statement is not obvious.
For a constant centre-of-mass speed, under rotational equilibrium, the same transition curves along the black line imply the same lean angle.
\section{Dissipative forces}
\label{sec:InstPower}
\begin{figure}[h]
\centering
\includegraphics[scale=0.35]{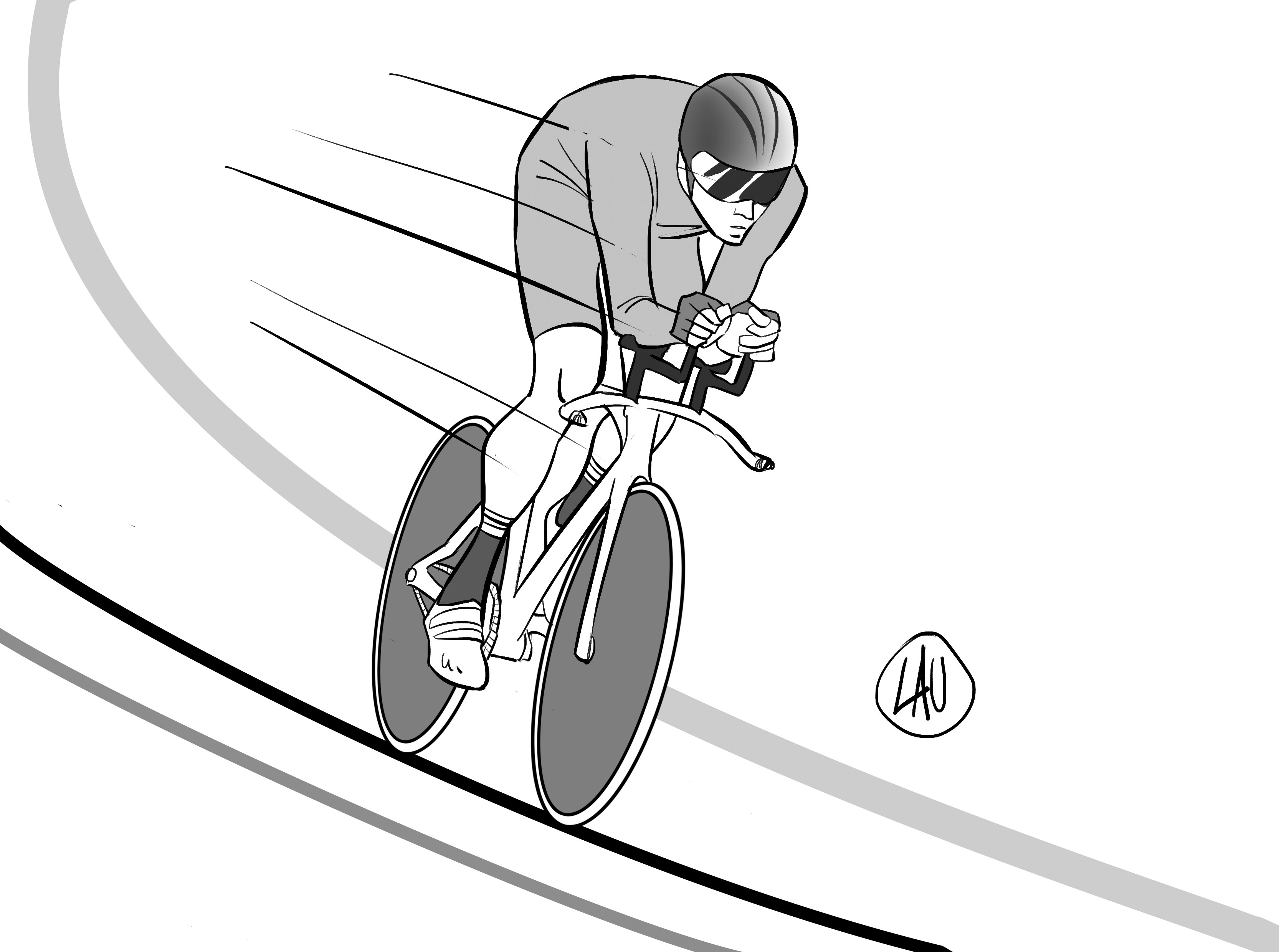}
\caption{\small  A constant aerodynamic position along the black line}
\label{fig:FigBlackLine}
\end{figure}

A mathematical model for the power required to propel a bicycle against dissipative forces is
\begin{equation}
\label{eq:BikePower}
P_F=F\,V\,,
\end{equation}
where $F$ stands for the magnitude of forces opposing the motion and $V$ for the centre-of-mass speed.%
\footnote{In expression~(\ref{eq:power}), below, we distinguish between the centre-of-mass speed,~$V$, to account for the air resistance, and the wheel speed,~$v$, to account for the rolling resistance and the lateral friction of the wheels.}
We model the bicycle-cyclist system as undergoing instantaneous circular motion in a horizontal plane, in rotational equilibrium about the instantaneous centre-of-mass velocity, $\bf V$.
We assume that the cyclist maintains a constant aerodynamic position, as illustrated in Figure~\ref{fig:FigBlackLine}.

Even though this model involves several simplifying assumptions, it captures the mechanical behaviour of the system, while allowing us to derive closed-form expressions. 
Specifically, the assumption of instantaneous circular motion allows us to model motion along a black line of arbitrary geometry in terms of a radius of curvature that varies from point to point.
In examining dissipative forces, the assumption that this circular motion occurs in a horizontal plane, although not strictly valid\,---\,in particular, since the centre of mass moves along a three-dimensional curve\,---\,is an adequate description of a motion whose horizontal displacements have much larger magnitudes than vertical displacements.

Similarly, the assumption of rotational equilibrium about~$\bf V$\,, wherein the lean angle changes only along the transition curves, is not strictly valid, since physically\,---\,as the cyclist enters and exits a turn\,---\,the lean angle increases, reaches a maximum, and decreases.
However, this approximation successfully captures the mechanical behaviour of the system for lap-averaged considerations, discussed in Section~\ref{sec:MechEn}, below, while again allowing us to derive closed-form expressions.

\begin{figure}
\centering
\includegraphics[scale=0.7]{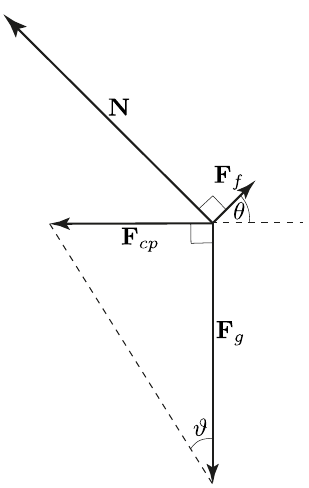}
\caption{\small Force diagram: Inertial frame}
\label{fig:FigFreeBody}
\end{figure}

In view of Figure~\ref{fig:FigFreeBody}, along the black line, in windless conditions,
\begin{subequations}
\label{eq:power}
\begin{align}
\nonumber P_F&=\\
&\dfrac{1}{1-\lambda}\,\,\Bigg\{\label{eq:modelO}\\
&\left.\left.\Bigg({\rm C_{rr}}\underbrace{\overbrace{\,m\,g\,}^{F_g}(\sin\theta\tan\vartheta+\cos\theta)}_N\cos\theta
+{\rm C_{sr}}\Bigg|\underbrace{\overbrace{\,m\,g\,}^{F_g}\dfrac{\sin(\theta-\vartheta)}{\cos\vartheta}}_{F_f}\Bigg|\sin\theta\Bigg)\,v
\right.\right.\label{eq:modelB}\\
&+\,\,\tfrac{1}{2}\,{\rm C_dA}\,\rho\,V^3\Bigg\}\label{eq:modelC}\,,
\end{align}	
\end{subequations}
where $m$ is the combined mass of the cyclist and the bicycle, $g$ is the acceleration due to gravity%
\footnote{The value of $g$ varies with latitude, $\phi$\,, and altitude,~$a$\,,
\begin{equation}
\label{eq:g}
	g = g_c\left(1 + \beta_1\sin^2\phi + \beta_2\sin^2 2\phi\right) -\left(\dfrac{2\,g_0}{r}\right)a\,,
\end{equation}
where $g_c = 9.780327\,{\rm m/s}^2$, $\beta_1 = 5.30244\times10^{-3}$, $\beta_2=-5.8\times10^{-6}$, $g_0 = 9.806257\,{\rm m/s}^2$ and $r = 6.371\times10^6\,\rm m$ \citep[e.g.,][Sections~2.4.4 and~2.5.4.4]{Lowrie2007}.
For consistency, to evaluate $g$ in summand~(\ref{eq:modelB}), we use expression~(\ref{eq:g}), even though it varies\,---\,at the most\,---\,by several parts in a thousand, which means that variability of $g$ does not affect the value of $P_F$ in model~(\ref{eq:power}).},
$\theta$ is the track-inclination angle, $\vartheta$ is the bicycle-cyclist lean angle, $\rm C_{rr}$ is the rolling-resistance coefficient, $\rm C_{sr}$ is the coefficient of the lateral friction, $\rm C_{d}A$ is the air-resistance coefficient, $\rho$ is the air density and $\lambda$ is the drivetrain-resistance coefficient.
In accordance with the ideal gas law, the air density is
\begin{equation}
\label{eq:AirDens}
	\rho=\dfrac{p\,M}{{\rm R}\,T}\,,
\end{equation}
wherein $p$ is the air pressure%
\footnote{\label{foot:AirPress}If not measured directly, the air pressure\,---\,at altitude~$a$\,---\,can be estimated using the isothermal atmosphere assumption with
\begin{equation}
\label{eq:AirDensAlt}
	p=p_0\exp\left(-\dfrac{g\,M}{{\rm R}\,T}\,a\right)\,,
\end{equation}
where $p_0$ is pressure at sea level \citep[e.g.,][Chapter~2.2]{BohrenAlbrecht1998}.
In general, to evaluate summand~(\ref{eq:modelC}), in predicting the value of $P_F$\,, we need both expressions~(\ref{eq:AirDens}) and (\ref{eq:AirDensAlt}).
For the accuracy of an {\it a posteriori} examination, however, only expression~(\ref{eq:AirDens}) could be used, with $p$ obtained by direct measurements.}, $M$ is the molar mass of dry air, $\rm R$ is the gas constant and $T$ is the absolute temperature.
We do not include humidity, whose effect on $\rho$ is much smaller than the effect of $p$\,, $M$ and $T$.

In expression~(\ref{eq:power}), $v$ is the speed of the contact point between the wheels and the track; we refer to it as the black-line speed.
The wheels are assumed to roll without slipping, so that $v$ is also the tangential speed of a point on the circumference of a wheel with respect to the axle.
$V$ is the centre-of-mass speed.
Since lateral friction is a dissipative force, it does negative work, and the work done against it\,---\,as well as the power\,---\,is positive.
For this reason, in expression~(\ref{eq:modelB}), we consider the magnitude,~$\big|{\,\,}\big|$\,.

Expression~(\ref{eq:power}) gives the instantaneous power required to overcome dissipative forces at a given point along the track, whose design is determined by its radius of curvature,~$R$\,, and its inclination,~$\theta$\,, which can be viewed as the two independent variables of expression~(\ref{eq:power}) in the sense that if $R$ and $\theta$ are constant, so is $P_F$. 
$R$ appears in expression~(\ref{eq:power}) implicitly\,---\,through expression~(\ref{eq:vV}), below\,---\,and $\theta$ appears explicitly.
Under the assumption of constant $V$\,---\,discussed in Section~\ref{sec:ConstV}, below\,---\,and given the values of $m$\,, $g$ and $\rho$\,, the two dependent variables are $\vartheta$  and $v$\,.
The four model parameters are $\lambda$\,, $\rm C_{rr}$\,, $\rm C_{sr}$ and $\rm C_dA$\,.
Since $P_F$ refers to a given point along the track, to obtain an average of power over an interval, say, a lap, we need to average the values of individual points within a lap.

To formulate summand~(\ref{eq:modelB}), we use the relations among the magnitudes of vectors~$\bf N$\,, ${\bf F}_g$\,, ${\bf F}_{cp}$ and ${\bf F}_f$\,, illustrated in Figure~\ref{fig:FigFreeBody}.
In accordance with Newton's second law, for a cyclist to maintain a horizontal trajectory, the resultant of all vertical forces must be zero,
\begin{equation}
\label{eq:Fy0}
\sum F_y=0=N\cos\theta+F_f\sin\theta-F_g
\,.
\end{equation}
In other words, ${\bf F}_g$ must be balanced by the sum of the vertical components of normal force, $\bf N$\,, and the friction force, ${\bf F}_f$\,, which is parallel to the velodrome surface and perpendicular to the instantaneous velocity.
Depending on the centre-of-mass speed and the radius of curvature for the centre-of-mass trajectory, if $\vartheta<\theta$\,, ${\bf F}_f$ points upwards, in Figure~\ref{fig:FigFreeBody}, which corresponds to its pointing outwards, on the velodrome;
 if $\vartheta>\theta$\,, it points downwards and inwards.
 If $\vartheta=\theta$\,, ${\bf F}_f=\bf 0$\,.
 Since we assume no lateral motion, ${\bf F}_f$ accounts for the force that prevents it.
 Heuristically, it can be conceptualized as the force exerted in a lateral deformation of the tires.

For a cyclist to follow the curved bank, the resultant of the horizontal forces,
\begin{equation}
\label{eq:Fx0}
\sum F_x=-N\sin\theta+F_f\cos\theta=-F_{cp}
\,,
\end{equation}
is the centripetal force,~${\bf F}_{cp}$\,, whose direction is perpendicular to the direction of motion and points towards the centre of the radius of curvature.
According to the rotational equilibrium about the centre of mass,
\begin{equation}
\label{eq:torque=0}
\sum\tau_z=0=F_f\,h\,\cos\left(\theta-\vartheta\right)-N\,h\,\sin(\theta-\vartheta)\,,
\end{equation}
where $\tau_z$ is the torque about the axis parallel to the instantaneous centre-of-mass velocity and $h$ is the upright height of the centre of mass of the bicycle-cyclist system. This condition implies
\begin{equation}
\label{eq:torque}
F_f=N\tan(\theta-\vartheta)\,,
\end{equation}
and $h$ is the centre-of-mass height of the bicycle-cyclist system at $\vartheta=0$\,.
Substituting expression~(\ref{eq:torque}) in expression~(\ref{eq:Fy0}), we obtain
\begin{equation}
\label{eq:N}
N=\dfrac{m\,g}{\cos\theta+\tan(\theta-\vartheta)\sin\theta}
=m\,g\,(\sin\theta\tan\vartheta+\cos\theta)
\,.
\end{equation}
Using this result in expression~(\ref{eq:torque}), we obtain
\begin{equation}
\label{eq:Ff}
F_f=m\,g\,(\sin\theta\tan\vartheta+\cos\theta)\tan(\theta-\vartheta)
=m\,g\,\dfrac{\sin(\theta-\vartheta)}{\cos\vartheta}\,.
\end{equation}

For summand (\ref{eq:modelC}), a formulation of ${\rm C_{d}A}\,\rho\,V^2/2$\,, therein, is discussed in Appendix~\ref{app:AirRes}, where we denote it by $F_a$\,.

We restrict our study to steady efforts, which\,---\,following the initial acceleration\,---\,are consistent with a pace of an individual pursuit or the Hour Record.
Steadiness of effort is an important consideration in relating the model, measurements and effort of a cyclist \citep[e.g.,][Sections~2.2 and 5.2]{DanekEtAl2021}.%
\footnote{\label{foot:fix}For a fixed-wheel drivetrain, rotations of pedals and wheels are directly linked.
Pedals spin as long as a bicycle moves, even if\,---\,in the extreme case\,---\,the cyclist's feet are not in contact with the pedals.
 
In accordance with expression~(\ref{eq:BikePower}), $P_F=F\,V$\,, model~(\ref{eq:power}) estimates the instantaneous power required to overcome dissipative forces,~$F$, to achieve a specific result, which can be expressed in terms of a time taken for a given distance, such as a 4000-metre pursuit, or in terms of a distance covered during a given time, such as the Hour Record.

For measurements of power, we use $p_{\circlearrowright}=f_{\circlearrowright} v_{\circlearrowright}$, where $f_{\circlearrowright}$ is the force applied to pedals by a cyclist and $v_{\circlearrowright}$ is their circumferential speed.
To relate measurements with physical conditions expressed within the model, we assume that $p_{\circlearrowright}$ is equal to the power required to overcome all opposing forces, including conservative forces, discussed in Section~\ref{sec:MechEn}, below; hence, $p_{\circlearrowright}>P_F$.

To relate measurements to the power generated by a cyclist, we integrate Newton's Second Law, ${\rm d}v_{\circlearrowright}/{\rm d}t=f_{\circlearrowright}/m$\,, to obtain
\begin{equation*}
v_{\circlearrowright}=v^0_{\circlearrowright}	+\frac{1}{m}\int\limits_0^tf_{\circlearrowright}\,{\rm d}t\,,
\end{equation*}
where $v^0_{\circlearrowright}$ is an initial speed, which\,---\,in  our model\,---\,is the speed achieved upon reaching a steady pace, and $m$ is the mass of the bicycle-cyclist system.
Thus, subsequent changes of speed,~${\rm d}v_{\circlearrowright}/{\rm d}t$\,, are proportional to the force at a given instant, even though $v_{\circlearrowright}$ depends on the history of~$f_{\circlearrowright}$.}
Furthermore, for road-cycling models \citep[e.g.,][expression~(1)]{DanekEtAl2021}\,---\,wherein curves are neglected so that acceleration has no centripetal component and there is no need to distinguish between the wheel speed and the centre-of-mass speed\,---\,such a restriction would correspond to setting the acceleration,~$a$\,, to zero.
On a velodrome, where\,---\,along the curves\,---\,acceleration has also a centripetal component, this is tantamount to a constant centre-of-mass speed,~${\rm d}V/{\rm d}t=0$\,.

Let us return to expression~(\ref{eq:power}).
Therein, $\theta$ is given by expression~(\ref{eq:theta}).
As shown in Appendix~\ref{app:LeanAng}, the lean angle is
\begin{equation}
\label{eq:LeanAngle}
\vartheta=\arctan\dfrac{V^2}{g\,r_{\rm\scriptscriptstyle CoM}}\,,
\end{equation}
where $r_{\rm\scriptscriptstyle CoM}$ is the centre-of-mass radius, and\,---\,along the curves, at any instant\,---\,the centre-of-mass speed is
\begin{equation}
\label{eq:vV}
V=v\,\dfrac{\overbrace{(R-h\sin\vartheta)}^{\displaystyle r_{\rm\scriptscriptstyle CoM}}}{R}
=v\,\left(1-\dfrac{h\,\sin\vartheta}{R}\right)\,,
\end{equation}
where $R$ is the radius discussed in Section~\ref{sub:Track}.
Along the straights, the  black-line speed is equal to the centre-of-mass speed, $v=V$\,.
As expected, $V=v$ if $h=0$\,, $\vartheta=0$ or $R=\infty$\,.
The lean angle\,, $\vartheta$\,, is determined by assuming that, at all times, the system is in rotational equilibrium about the line of contact of the tires with the ground; this assumption yields the implicit condition on $\vartheta$\,, stated in expression~(\ref{eq:LeanAngle}).
Hence, in accordance with expression~(\ref{eq:vV}), at any point of the track, $V$ is either equal to or smaller than~$v$.

As illustrated in Figure~\ref{fig:FigFreeBody}, expressions~(\ref{eq:LeanAngle}) and (\ref{eq:vV}) assume instantaneous circular motion of the centre of mass to occur in a horizontal plane.
Therefore, using these expressions\,---\,to examine dissipative forces\,---\,implies neglecting the vertical motion of the centre of mass.
Accounting for the vertical motion of the centre of mass would result in a nonhorizontal centripetal force.

To examine dissipative forces, we also neglect the effect of air resistance of rotating wheels \citep[e.g.,][Section~5.5]{DanekEtAl2021}.
We consider only the air resistance due to the translational motion of the bicycle-cyclist system.
Also, as commented in Appendix~\ref{app:AirRes}, in view of a steady ride and the quantification of average properties per lap, we assume $\rm{C_dA}$ to be constant.
It differs from one cyclist to another but, for a given cyclist, does not vary significantly with small variations in speed or\,---\,for individual time trials\,---\,with location along the track.
\section{Conservative forces}
\label{sec:MechEn}
\subsection{Mechanical energy}
\label{sub:Intro}
\begin{figure}
\centering
\includegraphics[scale=0.35]{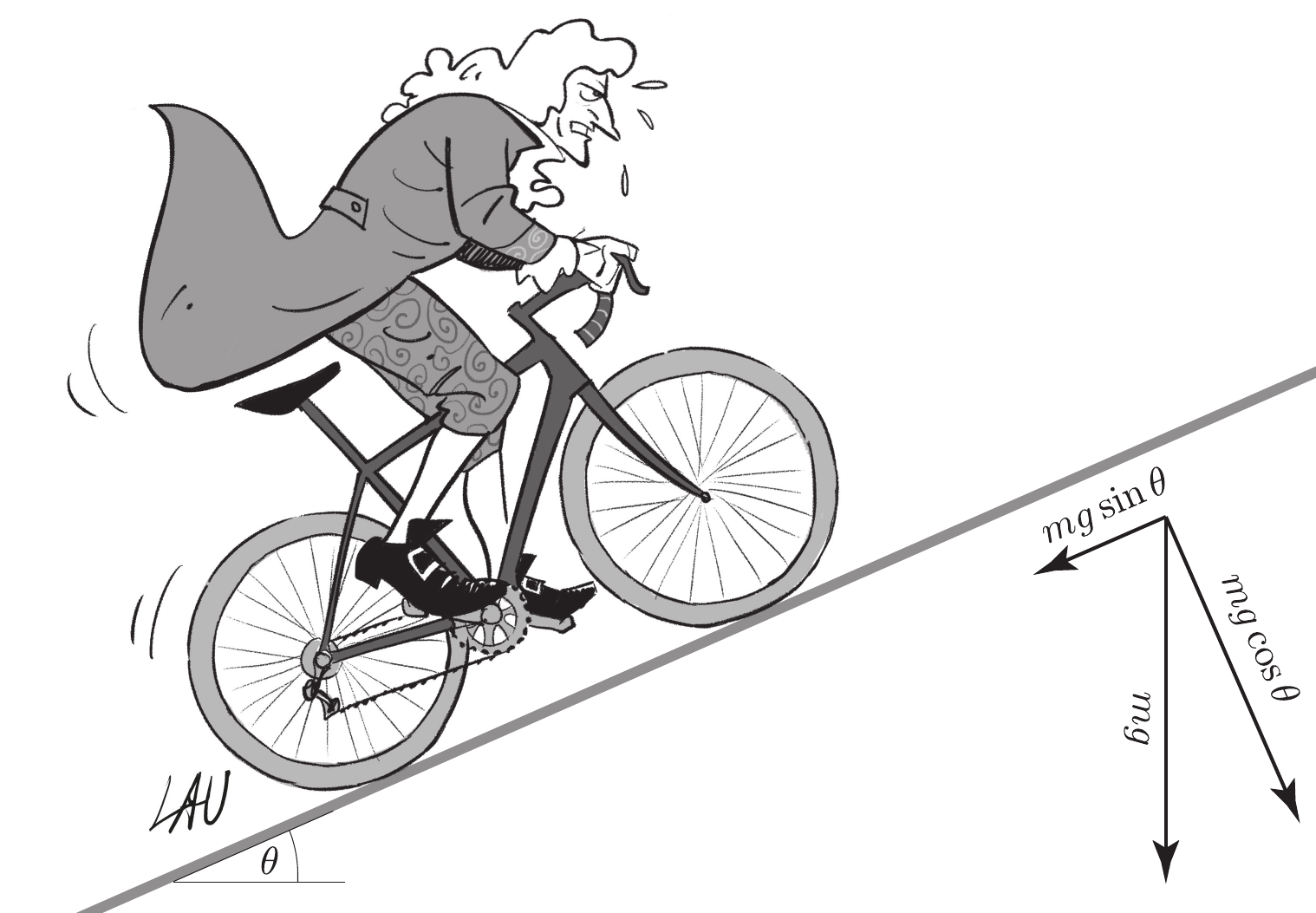}
\caption{\small Newton increasing gravitational potential energy while climbing a hill}
\label{fig:FigNewton}
\end{figure}

\begin{figure}
\centering
\includegraphics[scale=0.35]{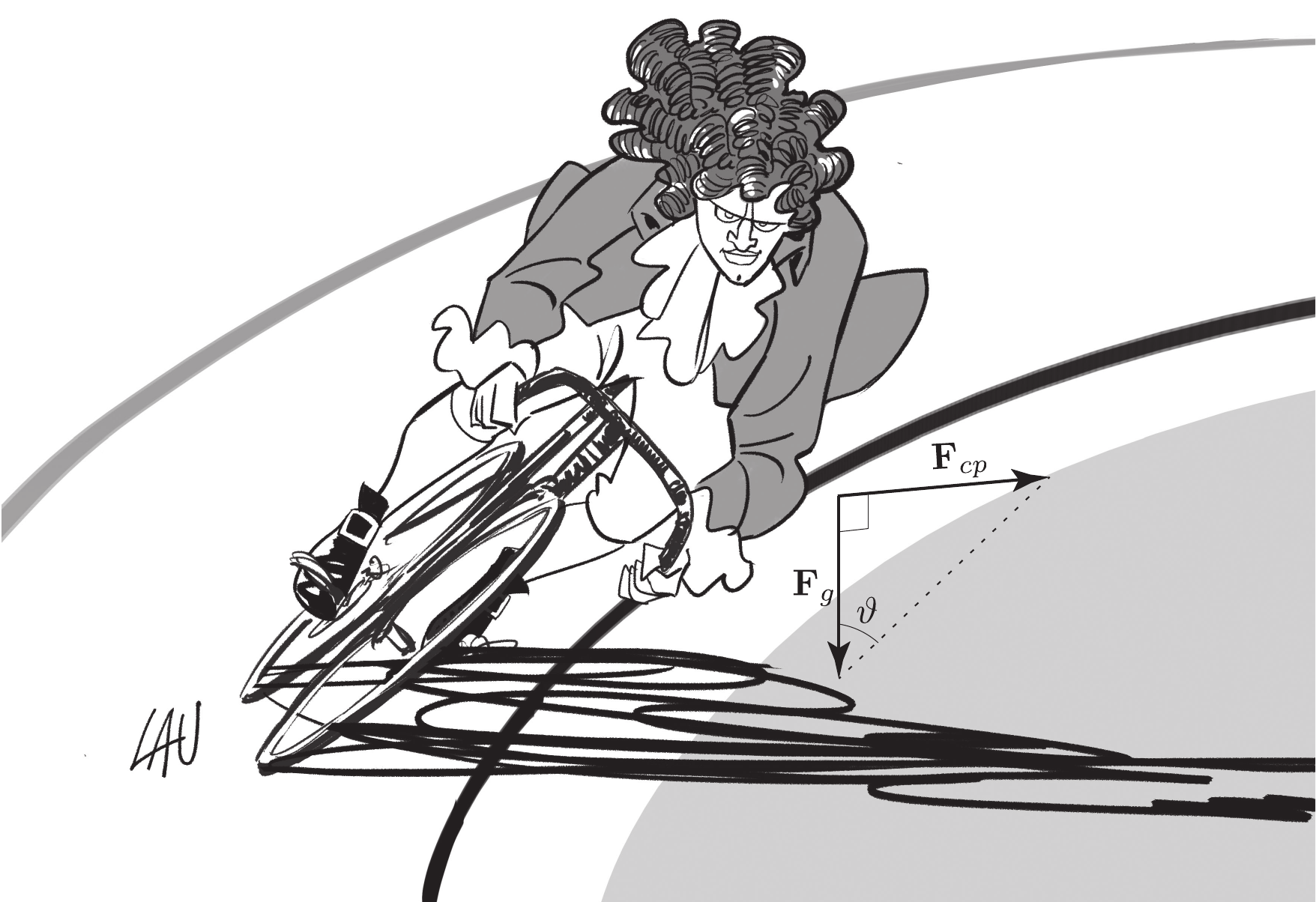}
\caption{\small Huygens increasing gravitational potential energy while exiting the curve}
\label{fig:FigHuygens}
\end{figure}

Commonly, in road-cycling models \citep[e.g.,][expression~(1)]{DanekEtAl2021}, we include $m\,g\sin\theta$\,, where\,---\,as illustrated in Figure~\ref{fig:FigNewton}\,---\,$\theta$ corresponds to a slope; this term represents force due to changes in gravitational potential energy associated with hills.
This term is not included in expression~(\ref{eq:power}). 
However\,---\,even though we assume the black-line of a velodrome to be horizontal%
\footnote{In accordance with the UCI regulations~\citep{UCI}, the black line is a curve located twenty centimetres from the inside edge of the track in the outward-normal direction.
Since this edge is horizontal and the track\,---\,in accordance with the UCI regulations\,---\,is a ruled surface with varying inclination angle, the black line deviates slightly from a horizontal plane \citep{Stanoev2023}.
\citet{FittonSymons2018} include\,---\,in their numerical model\,---\,the deviation of the black line from a horizontal plane; they do so based on theodolite measurements.

As stated in Section~\ref{sub:TrackInc}, for our model, the minimum track inclination, $\theta_{\rm min} = 13^\circ$\,, corresponds to the midpoint of the straight, and the maximum of $\theta_{\rm max} = 46^\circ$ to the apex of the circular arc.
Thus, the change in the black-line height is $0.2\,(\sin\theta_{\rm max} - \sin\theta_{\rm min}) = 0.1\,\rm m$\,.
Hence, since this change is small in comparison to the length over which it takes place, which is a quarter of the $250$-metre track, we assume the black line to be within a horizontal plane.

An analogous deviation of the red line is noticeable with a naked eye, and even more so, of the blue line, which\,---\,in a manner similar to the black line\,---\,are located at fixed distances from the inside edge.}\,---\,we need to account for the power required to increase gravitational potential energy to raise the centre of mass upon exiting the curves, as illustrated in Figure~\ref{fig:FigHuygens}.

In Section~\ref{sec:InstPower}, we assume the cyclist's centre of mass to travel in a horizontal plane, even though, along the transition curves, the centre-of-mass velocity,~$\bf V$, has both horizontal and vertical components, since the centre of mass follows a three-dimensional trajectory, lowering while entering a turn and rising while exiting it.

Considering dissipative forces, we neglect the vertical component, since the magnitude of the horizontal component, corresponding to the cyclist's forward motion, is much larger than that of the vertical component, corresponding to the up and down motion of the centre of mass.
However, we cannot use this simplification to consider the change in the cyclist's gravitational potential energy. 
Hence, in Section~\ref{sub:PotEn}, below, we calculate this change using the difference in the vertical position of the centre of mass by invoking the lean angle, whose values are obtained from the assumption of instantaneous rotational equilibrium.

Since dissipative forces, discussed in Section~\ref{sec:InstPower}, are nonconservative, the average power expended against them depends on the details of the path travelled by the cyclist, not just on its endpoints.
For this reason, to calculate the average power expended against dissipative forces, we require a detailed model of the cyclist's path to calculate the instantaneous power.
Instantaneous power is averaged to obtain the average power expended against dissipative forces over a lap.

However, we do not require such detail for calculating the average power needed to increase the cyclist's mechanical energy.
Provided this increase is monotonic, the change in mechanical energy\,---\,and therefore the average power\,---\,does not depend on the details of the path travelled by the cyclist, but only on its endpoints.
In brief, the path dependence of work against dissipative forces requires us to examine instantaneous motion to calculate the associated average power, but the path independence of work to change mechanical energy makes the associated average power independent of such an examination.
As discussed in Appendix~\ref{app:PU}, to compare power predictions and measurements, we cannot restrict the power expended to increase mechanical energy to the values along transition curves; these power estimates need to be evaluated for a lap\,---\,or, at least, half-a-lap\,---\,average.
\subsection{Kinetic energy}
\label{sub:KinEn}
To discuss the effects of mechanical energy, let us consider a bicycle-cyclist system whose centre-of-mass, at a given instant, moves with speed~$V$.
Neglecting the kinetic energy of the rotating wheels, we consider the kinetic energy of the system due to the translational motion of its centre of mass,
\begin{equation*}
K = \tfrac{1}{2}\,m\,V^2\,.
\end{equation*}
A bicycle-cyclist system is not purely mechanical; in addition to kinetic and potential energy, the system possesses internal energy in the form of the cyclist's chemical energy.
When the cyclist speeds up, internal energy is converted into kinetic.
However, the converse is not true: when the cyclist slows down, kinetic energy is not converted into internal.
Thus, to determine the work the cyclist does to change kinetic energy, we should consider only the increases.
If the centre-of-mass speed increases monotonically from $V_1$ to $V_2$\,, the work done is the increase in kinetic energy,
\begin{equation}
\label{eq:DelK}
\Delta K = \tfrac{1}{2}\,m\,\left(V_2^2-V_1^2\right)\,.
\end{equation}
The stipulation of a monotonic increase is needed due to the nonmechanical nature of the system, with the cyclist doing positive work\,---\,with an associated decrease in internal energy\,---\,when speeding up, but not doing negative work\,---\,with an associated increase in internal energy\,---\,when slowing down.

In the case of a constant centre-of-mass speed, $\Delta K=0$\,.
Otherwise, for instance, in the case of a constant black-line speed \citep[e.g.,][Appendix~A, Figure~A.1]{BSSSbici4}, the increase of $V$ occurs twice per lap, upon exiting a curve; hence, the corresponding increase in kinetic energy is $\Delta K = m\,\left(V_2^2-V_1^2\right)$\,.
The average power, per lap, required for this increase is
\begin{equation}
\label{eq:PK}
\overline P_K=\dfrac{1}{1-\lambda}\dfrac{\Delta K}{t_\circlearrowleft}\,,
\end{equation}
where $t_\circlearrowleft$ stands for the laptime.
Thus, for a constant centre-of-mass speed,~$\overline P_K=0$\,.
\subsection{Potential energy}
\label{sub:PotEn}
Let us determine the work\,---\,and, hence, average power, per lap\,---\,performed by a cyclist to change the potential energy,
\begin{equation*}
    U = m\,g\,h\,\cos\vartheta\,.
\end{equation*}
Since we assume a model in which the cyclist is instantaneously in rotational equilibrium about the black line, only the increase of the centre-of-mass height is relevant for calculating the increase in potential energy, and so for obtaining the average power required for the increase.
Since the lean angle,~$\vartheta$\,, depends on the centre-of-mass speed,~$V$\,, as well as on the radius of curvature,~$R$\,, of the black line, $\vartheta$\,---\,and therefore the height of the centre of mass\,---\,changes if either $V$ or $R$ changes.
The work done is the increase in potential energy resulting from a monotonic decrease in the lean angle from $\vartheta_1$ to $\vartheta_2$\,,
\begin{equation}
\label{eq:DeltaU}
\Delta U = m\,g\,h\,\left(\cos\vartheta_2-\cos\vartheta_1\right)\,.
\end{equation}
The average power, per lap, required for this increase is
\begin{equation}
\label{eq:PU}
\overline P_U=\dfrac{1}{1-\lambda}\dfrac{\Delta U}{t_\circlearrowleft}\,;
\end{equation}
herein, in contrast to expression~(\ref{eq:DeltaU}),
\begin{equation*}
\Delta U = 2\,m\,g\,h\,\left(\cos\vartheta_2-\cos\vartheta_1\right)\,,
\end{equation*}
since the bicycle-cyclist system straightens twice per lap, upon exiting a curve.
The same considerations\,---\,due to the nonmechanical nature of the system\,---\,apply to the work done to change potential energy as to that done to change kinetic energy.
The cyclist does positive work\,---\,with an associated decrease in internal energy\,---\,when straightening up, but not negative work\,---\,with an associated increase in internal energy\,---\,when leaning into the turn.

The increase in gravitational potential energy is subject to the same drivetrain losses as the power expended against dissipative forces, stated in expression~(\ref{eq:power}), and to increase kinetic energy, stated in expression~(\ref{eq:PK}); hence, the presence of $\lambda$ in expression~(\ref{eq:PU}).
Viewed from the noninertial reference frame of a cyclist following a curved trajectory, the torque causing the lean angle to decrease and, hence, potential energy to increase is the fictitious centrifugal torque.
This torque arises from the speed of the cyclist, which is in turn the result of the power expended on pedaling.

Returning to the first paragraph of Section~\ref{sub:Intro}, we conclude that the changes in gravitational potential energy discussed herein are different from the changes associated with hills.
When climbing a hill, a cyclist does work to increase potential energy.
When descending, a portion of that energy is converted into kinetic energy of forward motion.
This is not the case with the work the cyclist does to straighten up.
When the cyclist leans, potential energy is not converted into kinetic energy of forward motion.
Likewise, for a cyclist straightening up while coming out of a turn, the kinetic energy of forward motion is not simply converted into potential energy.
For example, in the case of constant black-line speed, as the cyclist straightens, the centre of mass both speeds up and is raised.
Therefore, both kinetic and potential energies increase, meaning one cannot be converted into the other.
Instead, the rider performs work to increase both.
Similarly, for a cyclist leaning into a turn at constant black-line speed, both kinetic and potential energies decrease, again meaning one cannot be converted into the other.
In general, since the bicycle and cyclist do not form a purely mechanical system, the decrease in mechanical energy cannot be converted into an increase in the cyclist's internal energy.

In the case of constant centre-of-mass speed, which we consider in the remainder of this article, $\overline P_U$ is equal to the additional power required to maintain constant $V$ along the transition curve, upon leaving a circular arc.
Along the arcs, $v$ is increased, hence $V<v$\,---\,as shown in Figure~\ref{fig:FigBLspeed}, below, on page~\pageref{fig:FigBLspeed}\,---\,and, thus conceptually, $\overline P_U$ corresponds to work done to bring both speeds together by the end of the transition curve, since $V=v$\,, along the straights.
Let us emphasize that\,---\,according to our model, wherein we represent a bicycle-cyclist system by its centre of mass and neglect the kinetic energy of spinning wheels\,---\,the increase of $v$ is only a consequence of the inward lean and does not entail any increase of kinetic energy, since $V$ remains constant.
Such a representation of a system is supported by the fact that\,---\,as discussed in Appendix~\ref{app:RotationalEnergy}\,---\,the change in rotational energy due to the lean is much smaller than the change in gravitational potential energy due to the lowering and raising of the centre of mass.
Using values stated in Section~\ref{sub:NumEx}, below, and disk wheels with a mass of $0.9\,{\rm kg}$ each (Daniel Bigham, {\it pers.\!~comm.}, 2023), the changes in rotational and gravitational potential energy on the transitions curves are $7.9\,\rm J$ and $337\,\rm J$, respectively.
In other words, the change in rotational energy is nearly two orders of magnitude smaller than the change in gravitational potential energy.

As discussed in Appendix~\ref{app:PU}, instantaneous values of $P_U$ do not yield an empirically adequate model.
Consequently, in expression~(\ref{eq:PU}), we take $\overline P_U$ to be the laptime average.
We can do so since the work to straighten up\,---\,which within our model is confined to the transition curve\,---\,is performed against a conservative force and, hence, depends only on the endpoints of the lean\,---\,the maximum along the arcs and the minimum along the straights\,---\,not on the evolution of its intermediate values.
\section{Constant centre-of-mass speed}
\label{sec:ConstV}
\subsection{Formulation}
\label{sub:FormCoM}
In accordance with expression~(\ref{eq:power}), let us find the instantaneous values of $P$ to overcome dissipative forces, under the assumption of a constant centre-of-mass speed,~$V$\,.
This assumption is consistent with constraining our attention to steady rides and with examining its motion in terms of the centre of mass.

Stating expression~(\ref{eq:vV}), as
\begin{equation}
\label{eq:Vv}
v=V\dfrac{R}{R-h\sin\vartheta}\,,
\end{equation}
we write expression~(\ref{eq:power}) as
\begin{equation}
\label{eq:PowerCoM}
\begin{aligned}
P_F&=\dfrac{V}{1-\lambda}\,\,\Bigg\{\tfrac{1}{2}\,{\rm C_{d}A}\,\rho\,V^2\\
&+\left.\left.\Bigg({\rm C_{rr}}\,m\,g\,(\sin\theta\tan\vartheta+\cos\theta)\cos\theta
+{\rm C_{sr}}\Bigg|\,m\,g\,\dfrac{\sin(\theta-\vartheta)}{\cos\vartheta}\Bigg|\sin\theta\Bigg)\,\dfrac{R}{R-h\sin\vartheta}
\right.\right.\,\Bigg\}\,.
\end{aligned}
\end{equation}
Equation~(\ref{eq:LeanAngle}), namely,
\begin{equation*}
\vartheta=\arctan\dfrac{V^2}{g\,(R-h\sin\vartheta)}\,,
\end{equation*}
can be solved for~$\vartheta$\,---\,to be used in expression~(\ref{eq:PowerCoM})\,---\,given $V$\,, $g$\,, $R$\,, $h$\,; along the straights, $R=\infty$ and, hence, $\vartheta=0$\,; along the circular arcs, $R$ is constant, and so is $\vartheta$\,; along transition curves, the radius of curvature changes monotonically, and so does $\vartheta$\,.
Along these curves, the radii are given by solutions of equations discussed in Section~\ref{sub:Track} or \ref{sub:BiTraj}, depending of the choice of differentiability.

The values of $\theta$ are given\,---\,at discrete points\,---\,from expression~(\ref{eq:theta}), namely,
\begin{equation*}
\theta(s)=29.5-16.5\cos\!\left(\dfrac{4\pi}{S}s\right)\,.
\end{equation*}
The values of the bicycle-cyclist system, $h$\,, $m$\,, ${\rm C_{d}A}$\,, ${\rm C_{rr}}$\,, ${\rm C_{sr}}$ and $\lambda$\,, are provided, and so are the external conditions, $g$ and $\rho$\,.

Under the assumption of a constant centre-of-mass speed,~$V$\,, there is no power used to increase the kinetic energy, only the power,~$\overline P_U$\,, used to increase the gravitational potential energy.
The latter is obtained by expression~\eqref{eq:PU}, where the pertinent values of $\vartheta$ for that expression are taken from model~\eqref{eq:PowerCoM}.
\subsection{Numerical example}
\label{sub:NumEx}
\subsubsection{Input and output}
\label{subsub:InOut}
To examine the model, let us consider the following values: the Hour Record, which at the time of this article is $56\,792\,\rm m$ and which\,---\,following the initial laptime of $24\,\rm s$ and as discussed in Appendix~\ref{app:SpeedChange}\,---\,corresponds to subsequent laptimes of $15.8113\,\rm s$ and, hence, to the black-line speed of $v=15.8115\rm m/s$\,, the bicycle-cyclist mass $m=97\,\rm kg$\,, altitude and latitude of Grenchen, Switzerland, for which, in accordance with expression~(\ref{eq:g}), $g=9.80625\,{\rm m/s^2}$\,, measured air density $\rho=1.12\,\rm kg/m^3$\,, centre-of-mass height $h=1.1\,\rm m$\,, ${\rm C_dA}=0.184\,\rm m^2$\,, ${\rm C_{rr}}=0.0015$\,, ${\rm C_{sr}}=0.002$ and $\lambda=0.015$\,.

To model the Tissot Velodrome in Grenchen, we use expression~\eqref{eq:theta}, where the track inclination varies from $13^\circ$\,, at the midpoint of the straight, to $46^\circ$ at the apex of the circular arc.
Also, as discussed in Section~\ref{sub:Track}, the half-length of the straight is $19\,\rm m$\,, the transition curve is $13.5\,\rm m$ and the remaining portion of the arc\,---\,whose radius is $23.3863\,{\rm m}$, as calculated in Section~(\ref{sub:BiTraj}), for a $C^3$ transition curve\,---\,is $30\,\rm m$\,; these segments constitute a quarter of the $250$-metre track.

To evaluate the power of the bicycle-cyclist system, we discretize the black line using $501$ evenly spaced points.
For each point, we compute numerically the values of~$V$ and~$\vartheta$ that satisfy lean-angle expression~\eqref{eq:LeanAngle}, which we use in expression~\eqref{eq:Vv} to obtain~$v$.
We use these values, together with track inclination~\eqref{eq:theta}, in model~\eqref{eq:PowerCoM}.
Aspects of such computations are detailed in Appendix~\ref{app:PU}.

The average power per lap\,---\,excluding the initial lap and assuming a constant centre-of-mass speed,~$V$\,, afterwards\,---\,is $\overline P_\circlearrowleft=459.7192\,\rm W$\,, which is the sum of $\overline P_F=416.4016\,\rm W$  and $\overline P_U=43.3177\,\rm W$\,; thus, $90.5774\,\%$ of power is used to overcome dissipative forces and, hence, $9.4226\,\%$ to increase gravitational potential energy.
The corresponding centre-of-mass speed is $V=15.5010\,\rm m/s$\,, hence, following summand~(\ref{eq:modelC}), we calculate that $389.6292\,\rm W$ of $\overline P_F$ is used to overcome the air resistance; the remainder of $\overline P_F$ is used to overcome the other three dissipative forces, expressed in terms of $\lambda$\,, $\rm C_{rr}$ and $\rm C_{sr}$\,.

\begin{figure}
	\centering
	\includegraphics[width=0.7\textwidth]{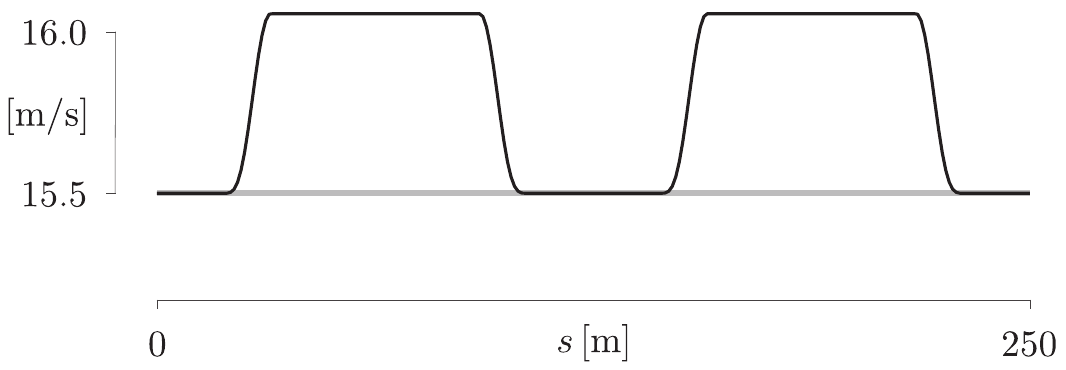}
	\caption{\small Instantaneous black-line (black) and centre-of-mass (gray) speeds}
	\label{fig:FigBLspeed}
\end{figure}

In accordance with expression~(\ref{eq:Vv}) and as shown in Figure~\ref{fig:FigBLspeed}, during each lap, the instantaneous black-line speed,~$v$\,, varies by $3.5830\,\%$\,, since, along the straights, $\vartheta=0^\circ$ and, hence, \mbox{$v=V=15.5010\,$\,m/s}\,, and, along the circular arcs\,---\,where, in accordance with expression~(\ref{eq:LeanAngle}), $\vartheta=47.3422^\circ$\,---\,$v=16.0564\,\rm m/s$\,.
Such an increase of the black-line speed along the curves agrees with empirical results.
Within the model, under the assumption of instantaneous rotational equilibrium, the changes of the lean angle,~$\vartheta$\,---\,and, hence, of the black-line speed,~$v$\,---\,are limited to the transition curves.

The instantaneous power,~$P_F$\,, varies only by~$3.0087\,\%$\,, which is consistent with a steady effort.
In view of the constancy of $V$, summand~(\ref{eq:modelC}), which is the power to overcome air resistance, is constant.
Summand~(\ref{eq:modelB}), which is the power to overcome rolling resistance and the lateral friction, increases slightly along the curve; this summand is a function of the lean angle,~$\vartheta$\,, the black-line speed,~$v$\,, and the track-inclination angle,~$\theta$\,.

Since the lean angle varies, $\vartheta\in[\,0^\circ,47.3422^\circ\,]$\,, for each curve, the centre of mass is raised by $1.1\,(\cos(0^\circ)-\cos(47.3422^\circ))=0.3546\,\rm m$\,.
For this Hour Record, which corresponds to $227$ completed laps, the centre of mass is raised by $2(227)(0.3546\,\rm m)=160.9973\,\rm m$\,.
As discussed in Sections~\ref{sub:KinEn} and~\ref{sub:PotEn}, there is no conversion of gravitational potential energy into kinetic energy upon leaning into the curve.
Since we model the motion by focusing on the centre of mass, whose speed is constant, the lean\,---\,with the corresponding increase of the black-line speed\,---\,does not entail an increase of kinetic energy, which remains constant along the entire lap.
\subsubsection{Sensitivity}
To gain insight into the sensitivity of the model to errors in input parameters, let us consider $m=97\pm1\,\rm kg$\,, $\rho=1.12\pm0.01\,\rm kg/m^3$\,, $h=1.1\pm 0.05\,\rm m$\,, ${\rm C_dA}=0.184\pm 0.01\,{\rm m^2}$\,, ${\rm C_{rr}}=0.0015\pm 0.0005$\,, ${\rm C_{sr}}=0.002\pm 0.0005$ and $\lambda=0.015\pm 0.005$\,.
For the lower limits, we obtain $\overline P_\circlearrowleft=422.8547\,\rm W$\,, and for the upper limits, $\overline P_\circlearrowleft=497.4264\,\rm W$\,.
The greatest effect is due to $\rm C_dA$\,, which is consistent with a dominant contribution of summand~(\ref{eq:modelC}) to the value of~$P_F$\,; if we consider only ${\rm C_dA}=0.184\pm 0.01\,{\rm m^2}$\,, we obtain $438.5437\,{\rm W}\leqslant \overline P_\circlearrowleft\leqslant480.8947\,\rm W$\,.
\begin{figure}[h]
	\centering
	\includegraphics[scale=0.45]{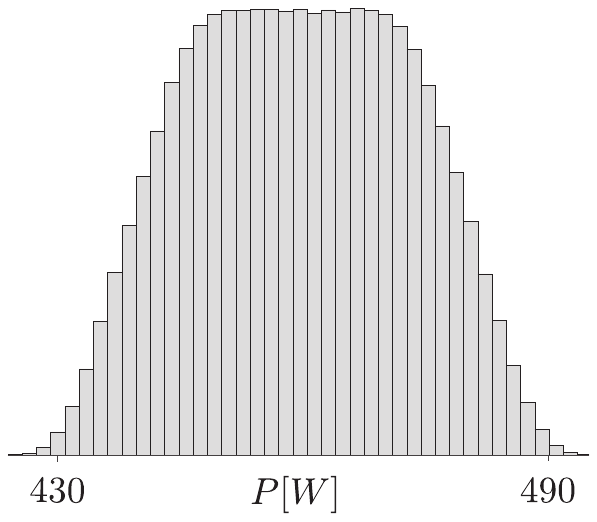}
\caption{\small Range of required power obtained by one million Monte-Carlo simulations from uniform distributions of $m$, $\rho$, $h$, $\rm C_dA$, $\rm C_{rr}$, $\rm C_{sr}$ and $\lambda$}
	\label{fig:FigMC}
\end{figure}

To gain insight into the distribution of the output, let us consider a Monte-Carlo simulation by sampling one million values from continuous uniform distributions of $m$, $\rho$, $h$, $\rm C_dA$, $\rm C_{rr}$, $\rm C_{sr}$ and $\lambda$\,, with the error ranges stated above, to obtain the histogram in Figure~\ref{fig:FigMC}.
The flat top of the histogram is due to the dominant effect of $\rm C_dA$\,.
The sample mean is $\mu = 459.7177\,\rm W$ and the maximum percentage error is $8.2364\%$\,.
Since $\mu$ is within $0.0003\%$ of the errorless prediction and the histogram is symmetric about the mean, the model is unbiased.
Also, the model is stable: small variations in input result in small variations in output.
However\,---\,in view of possible input errors\,---\,the absolute error in power estimates can be substantial.
\subsubsection{Different inputs}
\label{subsub:DifferentInputs}
To gain insight into the effect of significantly different inputs on the required power, let us consider several modifications.
\paragraph{Hour Record:}
Let the only modification be the reduction of Hour Record to $45\,000\,\rm m$\,.
The required average power reduces to~$\overline P_\circlearrowleft=233.1503\,\rm W$\,.
Thus, reducing the Hour Record by $21\%$ reduces the required power by about $49\%$\,.
Let the only modification be the increase of Hour Record to $60\,000\,\rm m$\,.
The required average power increases to~$\overline P_\circlearrowleft=540.2455\,\rm W$\,.
Thus, increasing the Hour Record by $6\%$ increases the required power by $18\%$\,.
These results are expected in view of a nonlinear relation between power and speed \citep[e.g.,][Section~4, Figure~11]{DanekEtAl2021}.
\paragraph{Altitude:}
Let the only modification be the increase of altitude to~$2600\,\rm m$\,, which corresponds to Cochabamba, Bolivia.
Following expressions~(\ref{eq:g}) and (\ref{eq:AirDens}), the corresponding adjustments are~\mbox{$g=9.7769\,\rm  m/s^2$} and $\rho=0.8754\,\rm kg/m^3$\,; hence, the required average power reduces to $\overline P_\circlearrowleft=374.6360\,\rm W$\,.
\paragraph{Temperature:}
Let the only modification be the reduction of temperature from $T=27^\circ{\rm C}$ to $T=20^\circ{\rm C}$\,.
Following expression~(\ref{eq:AirDens}), the corresponding adjustment is $\rho=1.1423\,\rm kg/m^3$\,; hence, the required average power increases to $\overline P_\circlearrowleft=467.4985\,\rm W$\,.
\paragraph{Mass:}
Let the only modification be the reduction of mass to $m=90\,\rm kg$\,.
The required average power reduces to $\overline P_\circlearrowleft=454.6612\,\rm W$\,.
In accordance with the formulation in Section~\ref{sub:PotEn}, the proportional reductions of $m$ and $\overline P_U$ are equal to one another.
\paragraph{Centre-of-mass height:}
Let the only modification be the reduction of the centre-of-mass height to $h=1\,\rm m$\,.
The corresponding adjustments are $\overline P_F=418.5302\,\rm W$\,, $\overline P_U=39.3968\,\rm W$ and \mbox{$V=15.5292\,\rm m/s$}\,; the required average power decreases to $\overline P_\circlearrowleft=457.9300\,\rm W$\,, which is a consequence of relations among $P_F$\,, $\overline P_U$\,, $V$ and $\vartheta$\,.
\paragraph{$\rm{\bf C_dA}$:}
Let the only modification be the reduction of the air-resistance coefficient to ${\rm C_dA}=0.175\,\rm m^2$\,.
The required average power reduces to~$\overline P_\circlearrowleft=440.6613\,\rm W$\,, which is a consequence of reduction of $P_F$ alone.
\paragraph{Track inclination:}
Let the only modification be $s_0=5$\,, in expression~(\ref{eq:s0}).
The required average power is~$\overline P_\circlearrowleft=459.7591\,\rm W$\,, which is a increase of less than~$0.1\%$\,.
\paragraph{Transition curve:}
Let the only modification be the $C^2$\,, as opposed to $C^3$\,, transition curve.
The required average power is~$\overline P_\circlearrowleft=459.7886\,\rm W$\,, which is an increase of less than~$0.1\%$\,.
\paragraph{Segment lengths:}
Let the only modification be the segment lengths such that the transition curve doubles in length.
To maintain $R=23.3863\,{\rm m}$, we require $L_s = 10.3858\,{\rm m}$, $L_t = 27.0000\,{\rm m}$, $L_a = 25.1142\,{\rm m}$.
The required average power is~$\overline P_\circlearrowleft=460.0787\,\rm W$\,, which is an increase of less than~$0.1\%$\,.
\subsection{Comparison with~\citet{FitzgeraldEtAl2021}}
\label{sub:Fitz}
Let us compare model~(\ref{eq:power}) to \citet[expression~(24)]{FitzgeraldEtAl2021},%
\footnote{Model~(\ref{eq:power}) appeared in \citet{BosEtAl2020} prior to \citet{FitzgeraldEtAl2021}.}
which\,---\,expressed in our notation\,---\,is
\begin{equation}
	\label{eq:FitzgeraldEtAl2021}
	\begin{aligned}
		P_F' &= \dfrac{1}{1-\lambda}\,\,\Bigg\{\tfrac{1}{2}\,{\rm C_dA}\,\rho V^3\\
		&\quad+\left({\rm C_{rr}}\,m\,g\,\dfrac{\cos(\vartheta-\theta)}{\cos\vartheta}+{\rm C_{rr}}\,\mu_{\rm s}\,m\,g\,\dfrac{\cos(\vartheta-\theta)}{\cos\vartheta}\,|\vartheta-\theta|\right)v\Bigg\};
	\end{aligned}
\end{equation}
it omits changes in gravitational potential and kinetic energy, but encompasses the dominant resistive forces for steady riding.
Model~\eqref{eq:FitzgeraldEtAl2021} differs from model~(\ref{eq:power}) by using $\mu_{\rm s}$\,, which is a scrubbing resistance~(e.g.,~\citet{LukesEtAl2012}) associated with the assumption of a slight shifting of the handlebars to avoid lateral slipping down the banked track, as opposed to $\rm C_{sr}$\,, the coefficient of lateral friction.
Both models assume that the bicycle-cyclist system follows the black line.
Notably, the same restriction is imposed by \citet{FittonSymons2018} to validate their model, even though, in general, they permit the black-line altitude to vary.

For the comparison, we use input parameters specified by~\citet[Section~3]{FitzgeraldEtAl2021}.
For the cyclist, $h=1$\,m, $m=75$\,kg, ${\rm C_dA} = 0.2\,{\rm m}^2$, $\rm C_{rr}=0.002$, $\mu_{\rm s} = 0.4125/$rad, $\lambda=0.02$.
For the environmental conditions, $g=9.81\,{\rm m/s}^2$, $\rho = 1.2\,{\rm kg/m}^3$.
For the velodrome, the length of the track is $S=250$\,m, the turn radius is $R=23.1950$\,m, the length of the transition is $b=24.9$\,m, and the bank angle~(\citet[eq.~(19)]{FitzgeraldEtAl2021}) is 
\begin{equation*}
	\theta(s) = \left(\dfrac{\theta_{\rm max}-\theta_{\rm min}}{2}\right)\sin\left(\dfrac{4\pi}{S}\,s-\dfrac{\pi}{2}\right) + \dfrac{\theta_{\rm max}+\theta_{\rm min}}{2},
\end{equation*}
where $\theta_{\rm min}=13^\circ$ and $\theta_{\rm max}=43^\circ$.
Using the expressions in Section~\ref{sub:Track}, the respective values of the remainder of the quarter circular arc, the half-length of the straight, and the Euler spiral coefficient are
\begin{equation*}
	c = \dfrac{\pi\,R - b}{2} = 23.9846\,{\rm m},\quad
	a = \dfrac{S}{4}-b-c = 13.6154\,{\rm m},\quad
	A = \dfrac{1}{R\,b} = 0.001731\,{\rm m}^{-2}.
\end{equation*}
In a manner consistent with our approach,~\citet{FitzgeraldEtAl2021} specify the instantaneous lean angle and centre-of-mass speed agreeing with expressions~\eqref{eq:LeanAngle} and~\eqref{eq:vV}\,---\,the numerical experiments assume a steady ride with $V=16$\,m/s, which corresponds to $\overline v_\circlearrowleft=16.2890\,\rm m/s$ and, hence,~\mbox{$t_\circlearrowleft=15.3511\,\rm s$}\,.
Using these values, together with $\rm C_{sr}=0.0025$\,, we compare the models to obtain the lap averages of
\begin{equation*}
	\overline P_F = 531.2604\,{\rm W}
	\quad\text{ and }\quad
	\overline P_F' = 534.7769\,{\rm W}\,.
\end{equation*}
Hence, the power to overcome dissipative forces obtained by model~(\ref{eq:power}) and by the model of \citet{FitzgeraldEtAl2021} is consistent to within one percent: $100\,(|\overline P_F-\overline P_F'|/\overline P_F')=0.6576\%$\,, in spite of a different formulation of lateral forces.
Also, for either model, during a lap, the power varies by $2.56\%$ and $2.85\%$\,, respectively, which agrees with the assumption of a steady ride. 

However, as discussed in Section~\ref{sec:MechEn}, a velodrome power model requires considerations of not only dissipative but also conservative forces.
Herein, for either model, $\overline P_U = 34.0442\,\rm W$\,, which \citet{FitzgeraldEtAl2021} do not include; hence, we claim that\,---\,for our and the~\citet{FitzgeraldEtAl2021} model\,---\,the required average power per lap is $565.3046\,\rm W$ and $568.8212\,\rm W$\,, respectively.
\section{Gradual speed increase}
\label{sec:SpeedChange}
Under certain conditions, our model\,---\,wherein we assume a constant centre-of-mass speed\,---\,can be used to estimate the power required for events of multiple laps, such as individual pursuit and Hour Record, with changing speed.
In this section, we show that\,---\,even though such an estimate violates the assumption used in Section~\ref{sub:FormCoM}\,---\,the resulting errors are negligible for pertinent scenarios.
To do so, we write expression~(\ref{eq:PK}) as
\begin{equation*}
  (\Delta\overline P_K)_i=\dfrac{1}{1-\lambda}\,\dfrac{(\Delta K)_i}{t_{\circlearrowleft_i}}\,,
  \end{equation*}
where
\begin{equation*}
(\Delta K)_i = \tfrac{1}{2}\,m\,\left(V_{\circlearrowleft_{i+1}}^2-V_{\circlearrowleft_i}^2\right)\,,
\end{equation*}
which is expression~(\ref{eq:DelK}) with $i$ specifying the lap number.

Let us set the time of the first lap to $24\,\rm s$\,.
The power required for the initial acceleration is not included in our study, which is limited to a steady, or nearly steady, ride of a cyclist who maintains an aerodynamic position.
Hence, model results in Table~\ref{table:Splits} begin with the second lap.

Following the initial lap, to reach $56\,792\,\rm m$ at the end of the hour\,---\,with a constant laptime for all subsequent laps\,---\,we require $\overline v_\circlearrowleft=56.9215\,\rm km/h$\,, which corresponds to $\overline P_\circlearrowleft=459.7192\,\rm W$\,, for each lap.

\begin{figure}[h]
\centering
\includegraphics[scale=0.55]{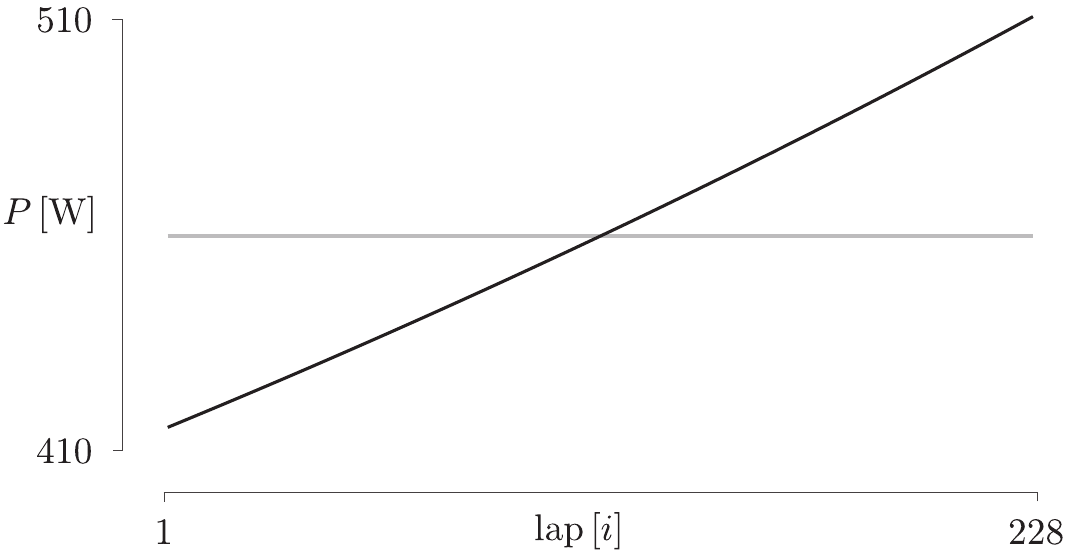}
\caption{\small Power required for constant and negative splits}
\label{fig:FigHRpower}
\end{figure}

To reach the same distance\,---\,with a linear decrease of laptime,%
\footnote{As shown in Figure~\ref{fig:FigHRpower}, for small and linear decrease of laptime, the corresponding increase of power is almost linear, even though it is a concave-up curve.} which is a measurable quantity readily available to a coach during an attempt, shown herein in the third column of Table~\ref{table:Splits}\,---\,let us set the average speed at the last lap to $\overline v_{\circlearrowleft n}=59.0000\,\rm km/h$\,.
Hence, as shown in Appendix~\ref{app:SpeedChange}, the average second-lap speed needs to be $\overline v_{\circlearrowleft 2}=54.9816\,\rm km/h$\,.

The average power over all completed laps\,---\,excluding the first one\,---\,is $\overline P=460.7787\,\rm W$ and the maximum lap average, which corresponds to the last completed lap is $\overline P_\circlearrowleft=510.5940\,\rm W$\,.
\begin{table}[h]
	\centering
	\begin{tabular}{c*{8}{c}}
    ${\rm lap}\,[i]$ & $d\,[\rm m]$ & $t_\circlearrowleft\,[\rm s]$ & $t\,[\rm h\!:\!m\!:\!s\!:\!ms]$ & $\overline v\,[\rm km/h]$ & $\overline P_\circlearrowleft\,[\rm W]$ & $\overline P\,[\rm W]$ & $V_\circlearrowleft\,[\rm m/s]$ & $\Delta\overline  P_K\,[\rm W]$ \\
    \toprule
    $1$ & $250$ & $24.000$ & $0\!:\!00\!:\!24.000$ & $37.5000$ & --- & --- & --- & --- \\
	$2$ & $500$ & $16.369$ & $0\!:\!00\!:\!40.369$ & $44.5886$ & $415.3158$ & $415.3158$ & $14.9828$ & $0.4009$ \\
$\vdots$ & $\vdots$ & $\vdots$ & $\vdots$ & $\vdots$ & $\vdots$ & $\vdots$ & $\vdots$ & $\vdots$\\
    $111$ & $27\,750$ & $15.829$ & $0\!:\!29\!:\!54.896$ & $55.6578$ & $458.2101$ & $436.2875$ & $15.4840$ & $0.4582$ \\
	$112$ & $28\,000$ & $15.824$ & $0\!:\!30\!:\!10.720$ & $55.6685$ & $458.6306$ & $436.4888$ & $15.4887$ & $0.4588$ \\
	$113$ & $28\,250$ & $15.819$ & $0\!:\!30\!:\!26.539$ & $55.6791$ & $459.0515$ & $436.6902$ & $15.4935$ & $0.4594$ \\
	$114$ & $28\,500$ & $15.814$ & $0\!:\!30\!:\!42.353$ & $55.6896$ & $459.4730$ & $436.8918$ & $15.4983$ & $0.4599$ \\
	$115$ & $28\,750$ & $15.809$ & $0\!:\!30\!:\!58.162$ & $55.7002$ & $459.8950$ & $437.0936$ & $15.5030$ & $0.4605$ \\
	$116$ & $29\,000$ & $15.804$ & $0\!:\!31\!:\!13.966$ & $55.7107$ & $460.3175$ & $437.2956$ & $15.5078$ & $0.4611$ \\
	$117$ & $29\,250$ & $15.799$ & $0\!:\!31\!:\!29.766$ & $55.7212$ & $460.7405$ & $437.4977$ & $15.5126$ & $0.4617$ \\
$\vdots$ & $\vdots$ & $\vdots$ & $\vdots$ & $\vdots$ & $\vdots$ & $\vdots$ & $\vdots$ & $\vdots$ \\
	$227$ & $56\,750$ & $15.254$ & $0\!:\!59\!:\!57.437$ & $56.7904$ & $510.5940$ & $460.7787$ & $16.0563$ & --- \\
	$228$ & $56\,792$ & $\,\,\,\,2.563$ & $1\!:\!00\!:\!00.000$ & $56.7920$ & --- & --- & --- & --- \\
	\bottomrule
    \end{tabular}
    \caption{Information corresponding to a constant laptime decrease}
    \label{table:Splits}
\end{table}

Such negative splits are considered desirable to delay the increase of internal body temperature, and were used by Daniel Bigham in his successful attempt in Summer 2022.
The average power for negative splits, $460.7787\,{\rm W}$, given in the penultimate row of the seventh column in Table~\ref{table:Splits}, is almost the same as the average lap power for a constant pace, $459.7192\,{\rm W}$, obtained in Section~\ref{subsub:InOut}; hence, in either case, almost the same amount of work is expended over an hour: $1\,658\,803.32\,{\rm J}$ and $1\,654\,989.12\,{\rm J}$\,, respectively.
However, as shown in Figure~\ref{fig:FigHRpower}, negative splits require a significantly higher maximum per-lap power, $510.5940\,\rm W$, given in the penultimate row of the sixth column in Table~\ref{table:Splits}.
Also, for the second half of the hour, the average lap power for negative splits is greater than for a constant pace: {\it Nullum gratuitum prandium}\,.

Comparing the values of the last column of Table~\ref{table:Splits} to the values of the sixth and seventh columns, we see that the amount of power required to accelerate is negligible.
Hence, the errors due to violating the assumption of a constant centre-of-mass speed are negligible.

A more direct, albeit less detailed, way to estimate the power required to increase from \mbox{$\overline{v}_{\circlearrowleft2}=54.9816\,\rm km/h$}\,, per lap, to $\overline{v}_{\circlearrowleft n}=59.0000\,\rm km/h$\,, per lap, which correspond to $V=14.9828\,\rm m/s$ and $V=16.0563\,\rm m/s$\,, respectively, is to consider the change in kinetic energy, $\Delta K=1616.0234\,\rm J$\,, over time, $(3600-24)\,\rm s$\,, to get $\Delta\overline  P_K=0.4588\,\rm W$\,, which is consistent with the values of the last column of Table~\ref{table:Splits}.
For either estimate, the speed increase requires about $0.1\%$ of the power needed to maintain a constant centre-of-mass speed per lap or of its average over all\,---\,but the first\,---\,laps.

The approach described in this section allowed us to generate scenarios for strategy considerations prior to two successful Hour Records in 2022: Daniel Bigham and Filippo Ganna.
We considered several different record predictions and used various speed increases, including cases where the speed for the second half of the hour remained constant, which was the preferred strategy of Daniel Bigham, who has optimized his performance with great attention to details.
The speed increase of $4\,\rm km/h$\,, illustrated herein, is the highest considered.
Hence, for other cases, the errors of estimated power\,---\,resulting from violating the assumption of constant~$V$\,---\,are even smaller.
\section{Conclusions}
\label{sec:DisCon}
The mathematical model presented in this article offers the basis for a quantitative study of individual time trials on a velodrome.
For a given power, the model can be used to predict laptimes and to estimate the expended power from the recorded times.
In the latter case, which is an inverse problem, the model can be used to infer values of parameters, in particular, $\rm C_dA$\,, $\rm C_{rr}$\,, $\rm C_{sr}$ and $\lambda$\,.
For inverse problems, the key measurable quantities are $v$ and $P$\,.%
\footnote{For a fixed-wheel drivetrain, wheel speed can be obtained from cadence, which is quite a reliable measurement.}
$V$ is not a directly measurable quantity; however, in accordance with expression~(\ref{eq:vV}), its value is smaller and varies less than $v$\,, which further supports the assumption of its being constant, upon which we base the formulation in Section~\ref{sec:ConstV}.

Let us emphasize that our model is phenomenological.
It is consistent with\,---\,but not derived from\,---\,fundamental concepts.
Its purpose is to provide quantitative relations between the model parameters and observables.
Its key justification is the agreement between measurements and predictions, which we comment on in Appendix~\ref{app:EmpAd}.
The relative simplicity of our model\,---\,as well as the model by \citet{FitzgeraldEtAl2021} and in contrast to \citet{FittonSymons2018}\,---\,facilitates an empirical examination of specific assumptions. 
While the accuracy of measurements is insufficient to distinguish between predictions of $\overline{P}_F = 531.2604\,{\rm W}$ versus $\overline{P}_F' = 534.7769\,{\rm W}$\,, discussed in Section~\ref{sub:Fitz}, it is possible to obtain empirical evidence for an increase of $\overline{P}_U = 34.0442\,\rm W$\,, predicted by our model.

Let us also comment on the $C^3$ transition curve, discussed in Section~\ref{sub:BiTraj} and used in all subsequent computations.
A black-line geometry with a discontinuity of a certain order in derivative of arclength with respect to time may result in a physical quantity, such as acceleration, not being well-defined at all points.
Besides the theoretical difficulty that this presents, such discontinuities can lead to unreliable numerical results.
Our methods allow one to make the track as smooth as necessary to overcome these computational issues, even though,  for the purposes of average per-lap power calculations, our discussion in Section~\ref{subsub:DifferentInputs} indicates that the $C^2$ continuity suffices.

Similarly to the choice of a $C^3$\,---\,as opposed to $C^2$\,---\,transition curve, four digits after the decimal point are included for numerical considerations, such as accuracy, within the model itself.
In view of measurement accuracy, there is no claim of using such precision to compare model predictions with data.
\section*{Acknowledgements}
We wish to acknowledge Mehdi Kordi, for information on track geometry and for measurements, used in Section~\ref{sec:Formulation} and Appendix~\ref{app:EmpAd};
Elena Patarini, for her graphic support;
David Dalton, for his editorial work;
Roberto Lauciello, for his artistic contributions;
Daniel Bigham, for fruitful information and discussions;
Francesco Sirio Basilico of Favero Electronics, for supporting our study with power meters with their technological advances, notably, IAV (Instantaneous Angular Velocity) Power, discussed by \citet[Appendix~A]{DanekEtAl2021}.
Following power-meter information from the latest\,---\,at the time of this article\,---\,successful Hour Record, we pursue the collaboration with Andrea Olivotto of Favero Electronics to enhance the agreement between the model and measurements since, inevitably, both are subject to inaccuracies and errors.
\bibliographystyle{apa}
\bibliography{BSSSvelodrome_arXiv}

\begin{thebibliography}{}

\bibitem[\protect\astroncite{Benham et~al.}{2020}]{BenhamEtAl2020}
Benham, G.~P., Cohen, C., Brunet, E., and Clanet, C. (2020).
\newblock Brachistochrone on a velodrome.
\newblock {\em Proceedings of the Royal Society A}, 476(2238).

\bibitem[\protect\astroncite{Birkhoff}{1950}]{Birkhoff}
Birkhoff, G. (1950).
\newblock {\em Hydrodynamics: A study in logic, fact, and similitude}.
\newblock Princeton University Press.

\bibitem[\protect\astroncite{Bohren and Albrecht}{1998}]{BohrenAlbrecht1998}
Bohren, C.~F. and Albrecht, B.~A. (1998).
\newblock {\em Atmospheric thermodynamics}.
\newblock Oxford University Press.

\bibitem[\protect\astroncite{Bos et~al.}{2020}]{BosEtAl2020}
Bos, L., Slawinski, M.~A., Slawinski, R.~A., and Stanoev, T. (2020).
\newblock On modelling bicycle power for velodromes: {P}art {II} {F}ormulation
  for individual pursuits.
\newblock {\em ar{X}iv}, 2009.01162v1 [physics.pop-ph].

\bibitem[\protect\astroncite{Bos et~al.}{2021}]{BSSSbici4}
Bos, L., Slawinski, M.~A., Slawinski, R.~A., and Stanoev, T. (2021).
\newblock On modelling bicycle power for velodromes: Part {II} {F}ormulation
  for individual pursuits.
\newblock {\em ar{X}iv}, 2009.01162 [physics.app-ph].

\bibitem[\protect\astroncite{Danek et~al.}{2021}]{DanekEtAl2021}
Danek, T., Slawinski, M.~A., and Stanoev, T. (2021).
\newblock On modelling bicycle power-meter measurements.
\newblock {\em ar{X}iv}, 2103.09806 [physics.pop-ph].

\bibitem[\protect\astroncite{Fitton and Symons}{2018}]{FittonSymons2018}
Fitton, B. and Symons, D. (2018).
\newblock A mathematical model for simulating cycling: applied to track
  cycling.
\newblock {\em Sports Engineering}, 21:409--418.

\bibitem[\protect\astroncite{Fitzgerald et~al.}{2021}]{FitzgeraldEtAl2021}
Fitzgerald, S., Kelso, R., Grimshaw, P., and Warr, A. (2021).
\newblock Impact of transition design on the accuracy of velodrome models.
\newblock {\em Sports Engineering}, 24(23).

\bibitem[\protect\astroncite{Lowrie}{2007}]{Lowrie2007}
Lowrie, W. (2007).
\newblock {\em Fundamentals of Geophysics}.
\newblock Cambridge University Press, 2nd edition.

\bibitem[\protect\astroncite{Lukes et~al.}{2012}]{LukesEtAl2012}
Lukes, R., Hart, J., and Haake, S. (2012).
\newblock An analytical model for track cycling.
\newblock {\em Proceedings of the Institution of Mechanical Engineers, Part P:
  Journal of Sports Engineering and Technolog}, 226(2):143--151.

\bibitem[\protect\astroncite{Martin et~al.}{1998}]{MartinEtAl1998}
Martin, J.~C., Milliken, D.~L., Cobb, J.~E., McFadden, K.~L., and Coggan, A.~R.
  (1998).
\newblock Validation of a mathematical model for road cycling power.
\newblock {\em Journal of Applied Biomechanics}, 14(3):276--291.

\bibitem[\protect\astroncite{Solarczyk}{2020}]{Solarczyk}
Solarczyk, M.~T. (2020).
\newblock Geometry of cycling track.
\newblock {\em {B}udownictwo i {A}rchitektura}, 19(2):111--119; doi:
  10.35784/bud--arch.1621.

\bibitem[\protect\astroncite{Stanoev}{2023}]{Stanoev2023}
Stanoev, T. (2023).
\newblock On technical considerations of velodrome track design.
\newblock {\em Sports Engineering}, 26(36).

\bibitem[\protect\astroncite{UCI}{2021}]{UCI}
UCI (2021).
\newblock {Union Cycliste Internationale} {R}egulations. {P}art {III}: {T}rack
  races. {R}etrieved in 2021 from:
  \href{https://www.uci.org/inside-uci/constitutions-regulations/regulations}{\tt
  https://www.uci.org/inside-uci/constitutions-regulations/regulations}.

\bibitem[\protect\astroncite{Underwood and Jermy}{2010}]{UnderwoodJermy2010}
Underwood, L. and Jermy, M. (2010).
\newblock Mathematical model of track cycling: the individual pursuit.
\newblock {\em Procedia Engineering}, 2(2):3217--3222.

\bibitem[\protect\astroncite{Underwood and Jermy}{2014}]{UnderwoodJermy2014}
Underwood, L. and Jermy, M. (2014).
\newblock Determining optimal pacing strategy for the track cycling individual
  pursuit event with a fixed energy mathematical model.
\newblock {\em Sports Engineering}, 17:183--196.

\end{thebibliography}
\begin{appendix}
\section{Air resistance}
\label{app:AirRes}
\setcounter{equation}{0}
\setcounter{figure}{0}
\renewcommand{\theequation}{\Alph{section}.\arabic{equation}}
\renewcommand{\thefigure}{\Alph{section}\arabic{figure}}
To formulate summand~(\ref{eq:modelC}), we assume that air resistance is proportional to the frontal surface area,~$A$\,, and to the pressure,~$p$\,, exerted by air on this area,~$F_a\propto p\,A$\,, where $p=\rho\,V^2/2$ has a form of kinetic energy and $V$ is the relative speed of the centre of mass with respect to the air; $p$ is the energy density per unit volume.
We can write this proportionality as
\begin{equation*}
F_a=\tfrac{1}{2}\,{\rm C_d}A\,\rho\,V^2\,,
\end{equation*}
where $\rm C_d$ is a proportionality constant, which is referred to as the drag coefficient.

A more involved justification for summand~(\ref{eq:modelC}) is based on dimensional analysis \citep[e.g.,][Chapter~3]{Birkhoff}.
We consider the air-resistance force, which is a dependent variable, as an argument of a function, together with the independent variables, to write
\begin{equation*}
f(F_a,V,\rho,A,\nu)=0\,; 	
\end{equation*}
herein, $\nu$ is the viscosity coefficient.
Since this function is zero in any system of units, it is only possible to express it in terms of dimensionless groups.

According to the Buckingham theorem \citep[e.g.,][Chapter~3, Section~4]{Birkhoff}\,---\,since there are five variables and three physical dimensions, namely, mass, time and length---\,we can express the arguments of~$f$ in terms of two dimensionless groups.
There are many possibilities of such groups, all of which lead to equivalent results.
A common choice for the two groups is
\begin{equation*}
\dfrac{F_a}{\tfrac{1}{2}\,\rho\,A\,V^2}\,,
\end{equation*}
which is referred to as the drag coefficient, and
\begin{equation*}
\dfrac{V\,\sqrt{A}}{\nu }\,,
\end{equation*}
which is referred to as the Reynolds number.
Thus, treating physical dimensions as algebraic objects, we can reduce a function of five variables into a function of two variables,
\begin{equation*}
g\left(\dfrac{F_a}{\tfrac{1}{2}\,\rho\,A\,V^2}\,,\,\dfrac{V\,\sqrt{A}}{\nu }\right)\,=\,0\,,
\end{equation*}
which we write as
\begin{equation*}
\dfrac{F_a}{\tfrac{1}{2}\,\rho\,A\,V^2}=\varphi\left(\dfrac{V\,\sqrt{A}}{\nu }\right)\,,
\end{equation*}
where the only unknown is~$F_a$\,, and where $\varphi$ is a function of the Reynolds number.
Denoting the right-hand side by $\rm C_d$\,, we write
\begin{equation*}
F_a=\tfrac{1}{2}\,{\rm C_d}\,A\,\rho\,V^2\,,
\end{equation*}
as expected.
In view of this derivation, $\rm C_d$ is not a constant; it is a function of the Reynolds number.
In our study, however\,---\,within a limited range of speed\,---\,$\rm C_d$ is treated as a constant.
Furthermore, since $A$ is difficult to estimate, we include it within this constant, and consider ${\rm C_d}\,A$ as a single value, which we denote as $\rm C_dA$\,.
\section{Lean angle}
\label{app:LeanAng}
\setcounter{equation}{0}
\setcounter{figure}{0}
\renewcommand{\theequation}{\Alph{section}.\arabic{equation}}
\renewcommand{\thefigure}{\Alph{section}\arabic{figure}}
In view of Figure~\ref{fig:FigFreeBody}, the lean angle of a cyclist is
\begin{equation}
\label{eq:vartheta}
\vartheta=\arctan\dfrac{F_{cp}}{F_g}\,,
\end{equation}
where the magnitude of the centripetal force is
\begin{equation}
\label{eq:Fc}
F_{cp}=\dfrac{m\,V^2}{r_{\rm\scriptscriptstyle CoM}}\,,
\end{equation}
and the magnitude of the force of gravity is $F_g=m\,g$\,.

To relate $F_{cp}$ and $\vartheta$\,, we use results~(\ref{eq:N}) and (\ref{eq:Ff}) in expression~(\ref{eq:Fx0}), to obtain
\begin{equation}
\label{eq:Fc_vartheta}
F_{cp}=N\sin\theta-F_f\cos\theta=m\,g\,\tan\vartheta\,,
\end{equation}
which is tantamount to expression~(\ref{eq:vartheta}).
Examining expressions~(\ref{eq:Fc}) and (\ref{eq:Fc_vartheta}) in the context of expression~(\ref{eq:vartheta}), we see that the lean angle is a function of the centre-of-mass speed and of the radius of curvature for the centre-of-mass trajectory; $m$ is cancelled.
\begin{figure}[h]
\centering
\includegraphics[scale=0.8]{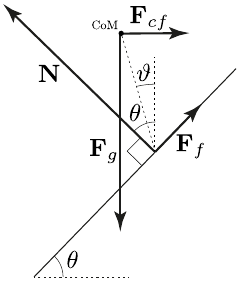}
\caption{\small Force diagram: Noninertial frame}
\label{fig:FigCentFric}
\end{figure}

Furthermore, the lean angle does not depend on the track inclination,~$\theta$\,, even though both the normal force,~$N$\, stated in expression~\eqref{eq:N}, and the frictional force,~$F_f$, stated in expression~\eqref{eq:Ff}, do depend on track inclination.
However, the $\theta$-dependence cancels out of the centripetal force,~$F_{cp}$\,, stated in expression~\eqref{eq:Fc_vartheta}, by a seldom used trigonometric identity.

Given the generality of this result, it would be satisfying to obtain it in a manner that explains it in the context of physics.
To this end, let us analyze the situation from inside the noninertial frame of the cyclist.
Specifically, we consider the frame comoving with the cyclist around a curve, with the cyclist considered as a point mass.
We neglect the additional accelerated motion resulting from the rotation of the cyclist about an axis through the centre of mass.

As illustrated in Figure~\ref{fig:FigCentFric} and in contrast to Figure~\ref{fig:FigFreeBody}, in this noninertial frame, instead of the forces in the horizontal direction summing to the centripetal force,~${\bf F}_{cp}$\,, the forces in this direction, including the fictitious centrifugal force, ${\bf F}_{cf}=-{\bf F}_{cp}$\,, sum to zero.
The centrifugal force must be taken to act at the centre of mass, since otherwise the torque about the centre of mass, stated in expression~\eqref{eq:torque=0}, would be affected.

To proceed, we invoke the vector identity,
\begin{equation*}
    \sum{\boldsymbol\tau} = {\bf R}_{\rm\scriptscriptstyle CoM}\times\sum{\bf F} + \sum{\boldsymbol\tau}_{\rm\scriptscriptstyle CoM}\,,
\end{equation*}
where $\sum{\boldsymbol\tau}$ is the net torque about an arbitrary point, $\sum{\boldsymbol\tau}_{\rm\scriptscriptstyle CoM}$ is the net torque about the centre of mass, ${\bf R}_{\rm\scriptscriptstyle CoM}$ is the position vector of the centre of mass, and $\sum{\bf F}$ is the net force.
From this identity there follows the well-known result that if the net force is zero and the net torque about the centre of mass is zero, the net torque about any other point is also zero.
In particular, let us consider the torque about the point of contact of the tires with the surface,
\begin{equation*}
\sum\tau_z = h\,F_g\sin\vartheta-h\,F_{cf}\sin\left(\dfrac{\pi}{2}-\vartheta\right) = 0\,,
\end{equation*}
where $h$ is the centre-of-mass height of the bicycle-cyclist system, from which\,---\,considering the magnitudes\,---\,it follows that
\begin{align}
\label{eq:lean_angle}
\nonumber\tan{\vartheta} &= \dfrac{F_{cf}}{F_g}\\
&= \dfrac{F_{cp}}{F_g},
\end{align}
where expression~\eqref{eq:lean_angle} is equivalent to expressions~\eqref{eq:vartheta} and \eqref{eq:Fc_vartheta}.
Expression~\eqref{eq:lean_angle} manifestly holds whether or not the track is banked.
\begin{figure}[h]
\centering
\includegraphics[scale=0.5]{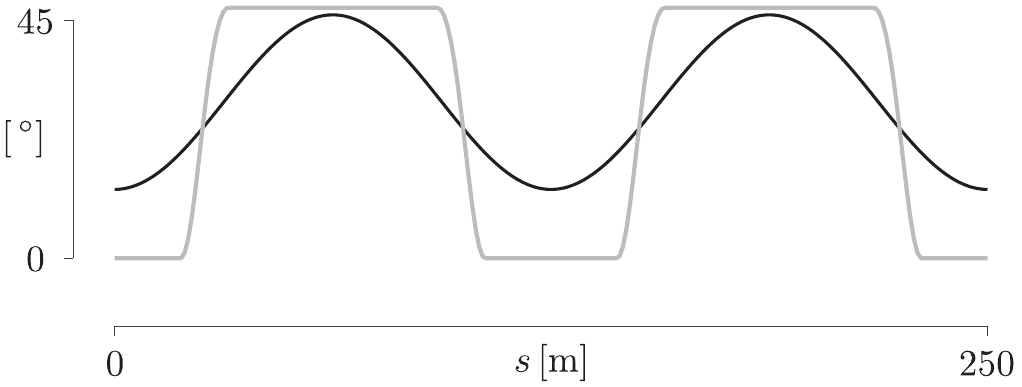}
\caption{\small Lean angle (gray) and track inclination (black)}
\label{fig:FigVarthetaTheta}
\end{figure}

From inside the noninertial frame of the cyclist, the value of the lean angle obtained for specific values of speed and radius is the one that makes the gravitational torque balance the centrifugal torque.
In particular, this condition makes no reference to track inclination.

As illustrated in Figure~\ref{fig:FigVarthetaTheta}, using input values of Section~\ref{sub:NumEx}\,---\,in view of instantaneous rotational equilibrium\,---\,the lean angle,~$\vartheta$\,, depends on the black-line curvature, shown in Figures~\ref{fig:FigTrackCurvature} and \ref{fig:FigBloss}\,---\,which is constant along the straights and circular arcs, changing only along the transition curves\,---\,not on track inclination,~$\theta$\,, shown in Figure~\ref{fig:FigAngle}.
\section{Power for increase of gravitational potential energy}
\label{app:PU}
\setcounter{equation}{0}
\setcounter{figure}{0}
\renewcommand{\theequation}{\Alph{section}.\arabic{equation}}
\renewcommand{\thefigure}{\Alph{section}\arabic{figure}}
In this appendix, we discuss\,---\,within the context of our model\,---\,the necessity of considering the power required to increase gravitational potential energy as an average, as formulated in Section~\ref{sub:PotEn}, not as instantaneous values.

\begin{figure}[h]
	\centering
	\includegraphics[width=0.6\textwidth]{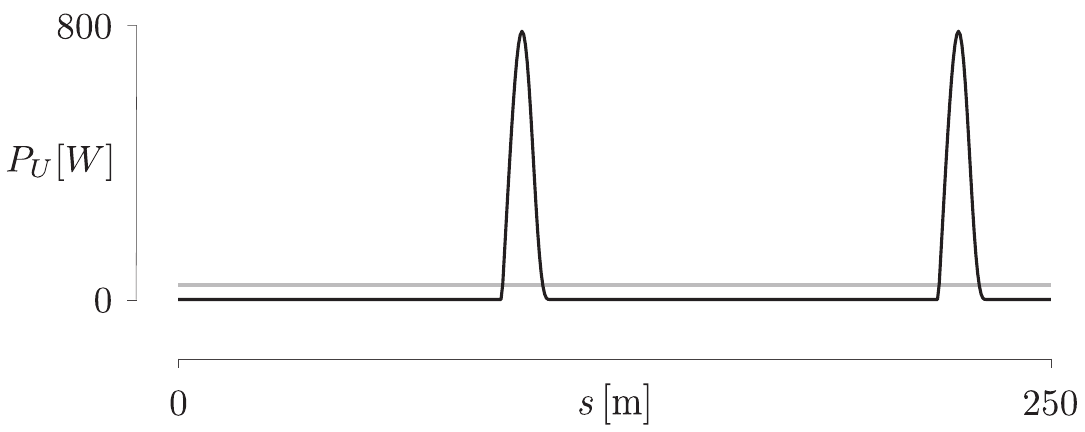}
	\caption{Instantaneous (black) and average (gray) power required to raise the centre of mass}
	\label{fig:FigPU}
\end{figure}

In Section~\ref{sub:NumEx}, we consider a bicycle-cyclist system with mass $m=97\,\rm kg$ and centre-of-mass height $h=1.1\,\rm m$ that completes a lap in $t_\circlearrowleft=15.8113\,\rm s$\,.
Along circular arcs, the cyclist lean angle is $\vartheta=47.3422^\circ$\,.
Upon leaving a circular arc, the bicycle-cyclist system increases its gravitational potential energy along the transition curve, as a result of raising the centre of mass.
In accordance with the assumption of instantaneous rotational equilibrium, the change in lean angle is restricted to the transition curve to reach $\vartheta=0^\circ$, by the beginning of the straight.
Thus, along each exiting transition curve, the centre of mass is raised by $\Delta h = h\left(\cos(0^\circ)-\cos(47.3422^\circ\right) = 0.3546\,\rm m$.
Since it occurs twice per lap, the increase in gravitational potential energy per lap is $\Delta U = 2\,m\,g\,\Delta h = 674.5955\,\rm J$\,.
As stated in expression~\eqref{eq:PU}, the average power, per lap, required for this increase of gravitational potential energy is
\begin{equation*}
	\overline{P}_U
	=
	\dfrac{1}{1-\lambda}\dfrac{\Delta U}{t_\circlearrowleft}
	=
	\dfrac{1}{1-0.015}\dfrac{674.5955\,\rm J}{15.8113\,\rm s}
	=
	43.3176\,\rm W.
\end{equation*}
As shown in Figure~\ref{fig:FigPU}, if we restrict the increase of gravitational potential energy to transition curves, we obtain power values that are inconsistent with empirical evidence; hence, in expression~\eqref{eq:PU} and in Section~\ref{sub:NumEx}, we consider an average per lap.

To obtain values presented in Section~\ref{sub:NumEx}, we use numerical computations along discrete points of the black line.
Within each interval along the transition curve, we assume the black-line speed to change linearly,
\begin{equation*}
	v(s)= v_i + \left(\dfrac{v_{i+1}-v_i}{s_{i+1}-s_i}\right)(s-s_i).
\end{equation*}

Therefore, on the $i$th interval,
\begin{equation*}
	v(s) = \dfrac{{\rm d}s}{{\rm d}t}.
\end{equation*}
Separating variables and integrating both sides, we write
\begin{equation*}
	\int\limits_{t_i}^{t_{i+1}}{\rm d}t = \int\limits_{s_i}^{s_{i+1}}\dfrac{{\rm d}s}{v(s)}\,.
\end{equation*}
To evaluate the right-hand side, we let
\begin{equation*}
	u = v(s)
	\quad\text{and}\quad
	{\rm d}u =\left(\dfrac{v_{i+1}-v_i}{s_{i+1}-s_i}\right){\rm d}s,
\end{equation*}
which changes the integration limits to $v_i$ and $v_{i+1}$.
Thus, the right-hand side becomes
\begin{equation*}
	\left(\dfrac{s_{i+1}-s_i}{v_{i+1}-v_i}\right)\int\limits_{v_i}^{v_{i+1}}\dfrac{{\rm d}u}{u}
=
	\left(\dfrac{s_{i+1}-s_i}{v_{i+1}-v_i}\right)\left(\ln(u)\Big|_{v_i}^{v_{i+1}}\right)
=
	\left(\dfrac{s_{i+1}-s_i}{v_{i+1}-v_i}\right)\ln\left(\dfrac{v_{i+1}}{v_i}\right)\,.
\end{equation*}
Hence, the time at the end of the $i$th interval is
\begin{equation}
	\label{eq:tip1}
	t_{i+1} = t_i + \left(\dfrac{s_{i+1}-s_i}{v_{i+1}-v_i}\right)\ln\left(\dfrac{v_{i+1}}{v_i}\right)\,.
\end{equation}
This expression is valid if $v_{i+1}\neq v_i$; if the black-line speed is constant, $t_{i+1}=t_i+(s_{i+1}-s_i)/v_i$\,.
Using expression~\eqref{eq:tip1}, the power required to increase the gravitational potential energy along the $i$th interval is
\begin{equation}
	\label{eq:PUi}
	(P_U)_i 
	=
	\dfrac{1}{1-\lambda}\dfrac{m\,g\,h\left(\cos\vartheta_{i+1}-\cos\vartheta_i\right)}{t_{i+1}-t_i}.
\end{equation}
Using input values of Section~\ref{sub:NumEx}, we plot expression~\eqref{eq:PUi} in Figure~\ref{fig:FigPU}.
$P_U=0$ everywhere except along the $13.5\,\rm m$ transition curves, where $\cos\vartheta_{i+1}\neq\cos\vartheta_i$\,; therein, the maximum value of each peak is $783.5231\,\rm W$\,.

Thus, we conclude that the assumption of instantaneous rotational equilibrium, which restricts changes of the lean angle to transition curves, does not result in an empirically adequate model for instantaneous values of $P_U$.
Since, in view of conservative forces, $P_U$ depends only on the endpoints of the lean, not on the evolution of its intermediate values, we consider its average per lap.
 In a similar manner, one could also consider a half-a-lap average.
\section{Rotational kinetic and gravitational potential energies}
\label{app:RotationalEnergy}
\setcounter{equation}{0}
\setcounter{figure}{0}
\renewcommand{\theequation}{\Alph{section}.\arabic{equation}}
\renewcommand{\thefigure}{\Alph{section}\arabic{figure}}
In this appendix, we compare the changes in gravitational potential energy of the centre of mass to the changes in rotational kinetic energy of wheels.
Both are caused by leaning.

\begin{figure}[h]
	\centering
	\begin{subfigure}[b]{0.59\textwidth}
		\centering
    	\begin{tikzpicture}
    		
    		\newcommand\Q{20};
    		\newcommand\QL{2};
    		\newcommand\hsinq{1.75};
    		\newcommand\lf{3};
    		\newcommand\R{8};
    		
    		\coordinate (O) at (0,0);
    		\coordinate (V0) at (\R-\hsinq,0);
    		\coordinate (v0) at (\R,0);
    		
    		\draw[dashed] ($(O)+(\QL,0)$)--(v0); 
    		
    		\draw[variable = \q, domain = 0:\Q]
    			plot ({\R*cos(\q)}, {\R*sin(\q)});
    		\draw[dashed,variable = \q, domain = 0:(-\Q)] 
    			plot ({\R*cos(\q)}, {\R*sin(\q)});
    		\draw[variable = \q, domain = 0:\Q]
    			plot ({(\R-\hsinq)*cos(\q)}, {(\R-\hsinq)*sin(\q)});
    		\draw[dashed,variable = \q, domain = 0:(-\Q)]
    			plot ({(\R-\hsinq)*cos(\q)}, {(\R-\hsinq)*sin(\q)});
    		
    		\draw[-] (O)--($(O)+(\QL,0)$);
    		\draw[-] (O)--($(O)+(\QL,{\QL*tan(\Q)})$);
    		\draw[variable = \q, domain = 0:\Q] 
    			plot ({0.8*\QL*cos(\q)}, {0.8*\QL*sin(\q)});
    		\draw 
    			node at ({\QL*cos(\Q/2)}, {\QL*sin(\Q/2)}) {$\Delta\phi$};
    		
    		\draw[|-|,thick] ($(V0)-(0,0.3)$)--($(v0)-(0,0.3)$)
    			node[centered,midway,fill=white] {$h\sin\vartheta$};
    			
    		\draw[|-|,thick] ($(v0)-(0,0.6)$)--($(O)-(0,0.6)$)
    			node[centered,midway,fill=white] {$R$};
    		
    		\draw[-Latex,ultra thick] 
    			(v0) -- ($(v0)+(0,2)$)
    			node[above] {$\boldsymbol v$};
    		\draw[-Latex,ultra thick] 
    			(V0) -- ($(V0)+(0,0.9*2)$)
    			 node[above] {$\boldsymbol V$};
		\end{tikzpicture}
		\caption{Centre-of-mass and black-line velocities}
		\label{fig:FigCircTurnVector}
	\end{subfigure}
	\begin{subfigure}[b]{0.39\textwidth}
		\centering
    	\begin{tikzpicture}
			
    		\newcommand\hsinq{1.75};
    		\newcommand\triS{0.25};	
    		
    		\coordinate (O) at (0,0);
		
    		\draw ($(O)+(0,-\hsinq)$)--($(O)+(0,\hsinq)$)
    			node[right,midway] {$h\cos\vartheta$};
    		\draw ($(O)+(0,\hsinq)$)--($(O)+(-\hsinq,\hsinq)$)
    			node[above,midway] {$h\sin\vartheta$};
    		\draw ($(O)+(-\triS,\hsinq)$)--($(O)+(-\triS,\hsinq)-(0,\triS)$)--($(O)+(0,\hsinq-\triS)$);
    		\draw ($(O)+(-\hsinq,\hsinq)$)--($(O)-(0,\hsinq)$)
    			node[left,midway] {$h$};
    		\draw[black,variable = \q, domain = (90+26):90] 
               	plot ({cos(\q)}, {-\hsinq+sin(\q)});
            \draw 
            	node at ({1.25*cos(90+13)}, {-\hsinq+1.25*sin(90+13)}) {$\vartheta$};
			
    	\end{tikzpicture}
		\caption{Lean angle of the centre of mass}
		\label{fig:FigCircTurnLean}
	\end{subfigure}
	\caption{Two distinct velocities as a consequence of lean angle}
	\label{fig:FigCircTurn}
\end{figure}
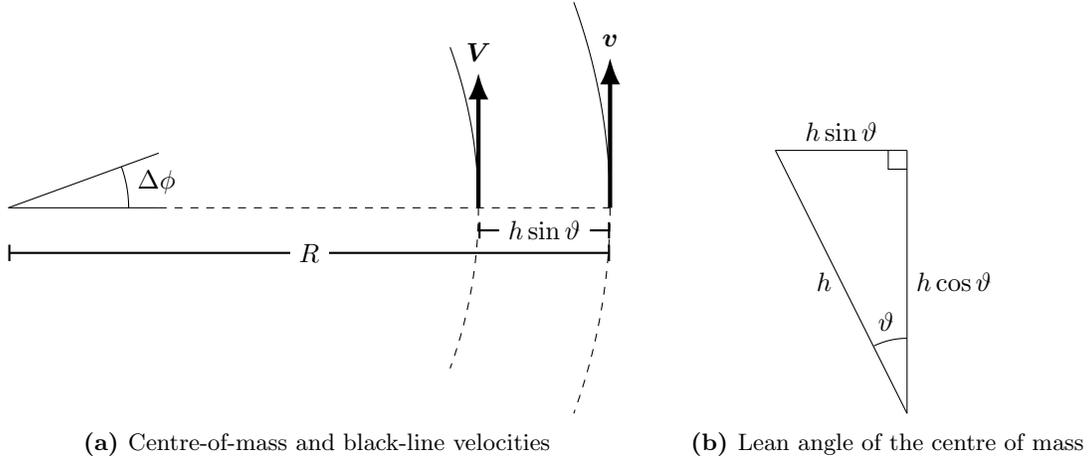

A cyclist entering a turn leans inward.
For constant centre-of-mass speed, this results in increased wheel speed\,---\,since, as shown in Figure~\ref{fig:FigCircTurn}, the wheels are traveling a longer path than the centre of mass\,---\,and therefore in increased rotational kinetic energy of the wheels.
At the same time, the lowering of the centre of mass results in a decrease in the gravitational potential energy of the centre of mass.
Conversely, a cyclist exiting a turn straightens up, resulting in\,---\,again, for constant centre-of-mass speed\,---\,a decrease in the rotational kinetic energy of the wheels and an increase in the gravitational potential energy of the centre of mass.

Can these changes be thought of in terms of energy conservation, with gravitational potential energy being converted into rotational kinetic energy when entering a turn, and vice versa when exiting a turn?
To answer this question, we calculate the magnitudes of the energy changes and show that, for typical choices of parameter values, the changes in gravitational potential energy are nearly two order of magnitude larger than the changes in rotational energy.
The decrease in rotational kinetic energy when exiting a turn contributes only negligibly to the increase in gravitational potential energy.
Consequently, the increase in gravitational potential energy must come almost exclusively from work done by the cyclist.

Consider a cyclist leaning inward.
The change in the rotational energy of both wheels when either entering or exiting the turn is
\begin{equation}
\label{eq:DeltaKrot}
	\Delta K_{\rm rot} 
	=
	\Delta\,2\left(\tfrac{1}{2}I\,\omega^2\right)
	=
	I\,\Delta\omega^2
	=
	I\,\Delta\left(\dfrac{v}{r}\right)^2
	=
	\dfrac{I}{r^2}\,\Delta v^2,
\end{equation}
where $I$, $r$ and $\omega$ are, respectively, the moment of inertia, radius and angular speed of a wheel, and $v$ is the black-line speed.

Consider a transition curve between a straight segment of the track and a circular arc with radius~$R$.
Along a straight the lean angle is zero and the wheel and centre-of-mass speeds are equal,
\begin{equation*}
	v = V.
\end{equation*}
As shown in Figure~\ref{fig:FigCircTurn}(\subref{fig:FigCircTurnVector}), for a circular arc with radius $R$ subtending an angle $\Delta\phi$ that is traversed in a time $\Delta t$, the wheel speed is
\begin{equation}
\label{eq:DelPhi}
	v = \dfrac{R\,\Delta\phi}{\Delta t}
\end{equation}
while\,---\,consistently with expression~(\ref{eq:vV})\,---\,the centre-of-mass speed is
\begin{equation*}
	V = \dfrac{(R-h\sin\vartheta)\,\Delta\phi}{\Delta t}\,,
\end{equation*}
where $h$ is the vertically upright height of the centre of mass and $\vartheta$ is the lean angle.
Using expression~(\ref{eq:DelPhi}) and rearranging, we see that\,---\,along the circular arc\,---\,the wheel and centre-of-mass speeds are
\begin{equation*}
	v =	\dfrac{V}{1-\dfrac{h}{R}\sin\vartheta}\,,
\end{equation*}
which is tantamount to expression~(\ref{eq:Vv}).
As shown in Figure~\ref{fig:FigCircTurn}(\subref{fig:FigCircTurnLean}), entering a turn, a cyclist leans inward and $\vartheta$ increases. As a result, for constant $V$, $v$ also increases.
Therefore, following expression~(\ref{eq:DeltaKrot}), over the entire transition curve between a straight segment and a circular arc, for a cyclist traveling at constant centre-of-mass speed, rotational energy increases by
\begin{equation*}
	\Delta K_{\rm rot}
	=
	\dfrac{IV^2}{r^2}\,\left(\dfrac{1}{\left(1-\dfrac{h}{R}\sin\vartheta_{\rm max}\right)^2}-1\right).
\end{equation*}
This result is independent of the detailed geometry of the transition curve; it depends only on the initial and final value of the lean angle.
Assuming disk wheels whose moment of inertia is
\begin{equation*}
	I = \dfrac{1}{2}m_{\rm w}\,r^2,
\end{equation*}
where $m_{\rm w}$ is the mass of a wheel, we obtain an increase in the rotational kinetic energy of the wheels due to the increased lean of the cyclist,
\begin{equation*}
	\Delta K_{\rm rot}
    =
	\dfrac{1}{2}m_{\rm w}V^2\,\left(\dfrac{1}{\left(1-\dfrac{h}{R}\sin\vartheta_{\rm max}\right)^2}-1\right).
\end{equation*}

To compare, let us consider the change in gravitational potential energy, $\Delta U$, due to a change in the lean and the resulting change in the height of the centre of mass.
Upon exiting a turn, the lean decreases and, hence, $U$ increases,
\begin{equation*}
	\Delta U
	=
	\Delta m\,g\,h
	=
	m\,g\,\Delta h
	=
	m\,g\,h(1-\cos\vartheta_{\rm max}),
\end{equation*}
which is tantamount to expression~(\ref{eq:DeltaU}); herein, we assume the final lean angle to be zero.

Using $R = 23.3863\,{\rm m}$\,, as discussed in Section~\ref{sub:BiTraj}, and the values stated in Section~\ref{subsub:InOut},
\begin{equation*}
	m = 97\,{\rm kg},\quad
	g = 9.80625\,{\rm m/s^2},\quad
	h = 1.1\,{\rm m},\quad
	\vartheta_{\rm max} = 47.3422^\circ,\quad
	V = 15.5010\,{\rm m/s}\,,
\end{equation*}
as well as $m_{\rm w} = 0.9000\,{\rm kg}$ (Daniel Bigham, {\it pers.\!~comm.}, 2023)
we obtain
\begin{equation*}
	\Delta K_{\rm rot} = 7.8872\,{\rm J}
\end{equation*}
and
\begin{equation*}
	\Delta U = 337.3167\,{\rm J}\,.
\end{equation*}
Thus, the change in the rotational kinetic energy of the wheels is 2.3\% of the change in the gravitational potential energy of the centre of mass.
\section{Laptimes for gradual speed increase}
\label{app:SpeedChange}
\setcounter{equation}{0}
\setcounter{figure}{0}
\renewcommand{\theequation}{\Alph{section}.\arabic{equation}}
\renewcommand{\thefigure}{\Alph{section}\arabic{figure}}
In Section~\ref{sec:SpeedChange}, we examine two pacing strategies for the Hour Record.
Following the first lap, one strategy assumes a constant laptime for all subsequent laps and the other assumes a linear decrease of laptime.
In this appendix, we derive the expressions used to obtain such laptimes. 

During an Hour Record attempt, a cyclist travels\,---\,within one hour,~$H$\,---\,a distance,~$D$, on a velodrome, whose length is $S$.
$D$ consists of $n$ laps and $r$ metres of the final incomplete lap.
For each complete $i$th lap, we denote its time by $t_{\circlearrowleft i}$ and the time along the incomplete lap by $t_r$\,; their sum is one hour,
\begin{equation}
	\label{eq:sumti_temp}
	\sum\limits_{i=1}^nt_{\circlearrowleft i} + t_{\circlearrowleft r} = H.
\end{equation}
To use this expression, we need to make three assumptions.
Since we do not model the initial acceleration\,---\,from a standstill to a steady speed\,---\,we assign the first laptime. 
We impose a linear interpolation of laptimes such that 
\begin{equation}
	\label{eq:t_i}
	t_{\circlearrowleft i} = t_{\circlearrowleft 2} + \left(\dfrac{t_{\circlearrowleft n}-t_{\circlearrowleft 2}}{n-2}\right)(i-2),
	\qquad\text{for}\quad i\in[\,2,n\,]\,.
\end{equation}
We assume the speed along the incomplete lap to be the speed of the~$n$th lap; hence,
\begin{equation}
	\label{eq:t_r}
	t_{\circlearrowleft r} = \left(\dfrac{r}{S}\right)t_{\circlearrowleft n}.
\end{equation}
Using expressions~\eqref{eq:t_i} and \eqref{eq:t_r} in expression~\eqref{eq:sumti_temp}, we write
\begin{equation}
	\label{eq:sumti}
	t_{\circlearrowleft 1} + \sum\limits_{i=2}^n\left(t_{\circlearrowleft 2} + \left(\dfrac{t_{\circlearrowleft n}-t_{\circlearrowleft 2}}{n-2}\right)(i-2)\right) + \left(\dfrac{r}{S}\right)t_{\circlearrowleft n} = H\,.
\end{equation}
As an aside, let us turn our attention to the summation term.
Its lower limit is not equal to one.
However, let us invoke a fact that
\begin{equation*}
	\sum\limits_{i=1}^nc = nc
	\quad\text{and}\quad
	\sum\limits_{i=1}^ni = \dfrac{n(n+1)}{2},
\end{equation*}
where $c$ is a constant value.
Then, to evaluate summations with variable lower limits, where~\mbox{$b>a>1$} and~$a,b\in\mathbb{N}$, we write
\begin{equation}
    \label{eq:sum_{i=a}^bc}
	\sum\limits_{i=a}^bc 
	=
	\left(\sum\limits_{i=1}^bc\right) - 
	\left(\sum\limits_{i=1}^{a-1}c\right)
	=
	bc - (a-1)c
	=
	(b-a+1)c
\end{equation}
and
\begin{equation}
    \label{eq:sum_{i=a}^bi}
	\sum\limits_{i=a}^bi 
	=
	\left(\sum\limits_{i=1}^bi\right) - 
	\left(\sum\limits_{i=1}^{a-1}i\right)
	=
	\dfrac{b(b+1)}{2} - \dfrac{(a-1)(a-1+1)}{2}
	=
	\dfrac{b(b+1)-(a-1)a}{2}.
\end{equation}
Hence, to evaluate the summation in expression~\eqref{eq:sumti}, we use the distributive property to obtain
\begin{equation*}
	\sum\limits_{i=2}^n\left(t_{\circlearrowleft 2} + \left(\dfrac{t_{\circlearrowleft n}-t_{\circlearrowleft 2}}{n-2}\right)(i-2)\right)
	=
	\sum\limits_{i=2}^nt_{\circlearrowleft 2} + \left(\dfrac{t_{\circlearrowleft n}-t_{\circlearrowleft 2}}{n-2}\right)\left(\sum\limits_{i=2}^ni-\sum\limits_{i=2}^n2\right).
\end{equation*}
Using identities~\eqref{eq:sum_{i=a}^bc} and~\eqref{eq:sum_{i=a}^bi}, 
\begin{equation*}
	\sum\limits_{i=2}^nt_{\circlearrowleft 2} + \left(\dfrac{t_{\circlearrowleft n}-t_{\circlearrowleft 2}}{n-2}\right)\left(\sum\limits_{i=2}^ni-\sum\limits_{i=2}^n2\right)
	=
	(n-1)t_{\circlearrowleft 2} + \left(\dfrac{t_{\circlearrowleft n}-t_{\circlearrowleft 2}}{n-2}\right)\left(\dfrac{n(n+1)-2}{2}-(n-1)(2)\right).
\end{equation*}
Then, we simply and factor to write summation in expression~\eqref{eq:sumti} as
\begin{equation}
	\label{eq:sumti_2ndTerm}
	\sum\limits_{i=2}^n\left(t_{\circlearrowleft 2} + \left(\dfrac{t_{\circlearrowleft n}-t_{\circlearrowleft 2}}{n-2}\right)(i-2)\right)
	=
	(n-1)\left(\dfrac{t_{\circlearrowleft 2}+t_{\circlearrowleft n}}{2}\right).
\end{equation}%
Returning to equation~\eqref{eq:sumti} and using result~\eqref{eq:sumti_2ndTerm}, we solve for $t_2$\,,
\begin{equation}
	\label{eq:t2_ramp}
	t_{\circlearrowleft 2} 
	=
	\dfrac{2(H-t_1)}{n-1} - \left(1+\dfrac{2r}{(n-1)S}\right)t_{\circlearrowleft n}\,.
\end{equation}
If $t_{\circlearrowleft n} = t_{\circlearrowleft 2}$\,, which corresponds to a constant-pace strategy, expression~\eqref{eq:t2_ramp} simplifies to
\begin{equation}
	\label{eq:t2_flat}
	t_{\circlearrowleft 2} 
	=
	\dfrac{S(H-t_1)}{r+(n-1)S}\,,
\end{equation}
which is the laptime for the second\,---\,and all subsequent\,---\,laps.
In Section~\ref{sec:SpeedChange}, $D = 56\,792\,{\rm m}$, \mbox{$S=250\,{\rm m}$}, $H=3600\,{\rm s}$ and $t_1 = 24\,{\rm s}$.
Hence, using modular arithmetic, we calculate
\begin{equation*}
	r = D\,\,\text{mod}\,\,S = 56\,792\,{\rm m}\,\,\text{mod}\,\,250\,{\rm m} = 42\,{\rm m}
\end{equation*}
and
\begin{equation*}
	n = \dfrac{D-r}{S} = \dfrac{56\,792\,{\rm m}-42\,{\rm m}}{250\,{\rm m}} = 227\,.
\end{equation*}
Thus, expression~\eqref{eq:t2_flat} results in
\begin{equation*}
	t_{\circlearrowleft 2}
	= 
	\dfrac{(250\,{\rm m})(3600\,{\rm s}-24\,{\rm s})}{(42\,{\rm m}) + (227-1)(250\,{\rm m})}
	=
	15.8113\,{\rm s}\,,
\end{equation*}
which corresponds to $\overline v_{\circlearrowleft 2} = 56.9215\,{\rm km/h}$ and which remains the same for all subsequent laps.

For an increasing-pace strategy, we need to choose $\overline v_{\circlearrowleft n}$\,, the average speed on the final lap, to obtain the corresponding laptime, $t_{\circlearrowleft n} = S/\overline v_{\circlearrowleft n}$\,.
Letting $\overline v_{\circlearrowleft n}=59.0000\,{\rm km/h}=16.3889\,\rm m/s$, we obtain $t_{\circlearrowleft n} =15.2542\,\rm s$\,, which appears in the penultimate row of Table~\ref{table:Splits}, on page~\pageref{table:Splits}.
Hence, according to expression~\eqref{eq:t2_ramp},
\begin{equation*}
	t_{\circlearrowleft 2}
	=
	\dfrac{2(3600\,{\rm s}-24\,{\rm s})}{227-1} - 
	\left(1+\dfrac{2(42\,{\rm m})}{(227-1)(250\,{\rm m})}\right)(15.2542\,{\rm s})
	=
	16.3691\,{\rm s}\,,
\end{equation*}
which appears in the second row of Table~\ref{table:Splits} and which corresponds to $\overline v_{\circlearrowleft 2} = 54.9816\,{\rm km/h}$.
\section{Empirical adequacy}
\label{app:EmpAd}
\setcounter{equation}{0}
\setcounter{figure}{0}
\renewcommand{\theequation}{\Alph{section}.\arabic{equation}}
\renewcommand{\thefigure}{\Alph{section}\arabic{figure}}
\begin{figure}[h]
\centering
\includegraphics[scale=0.5]{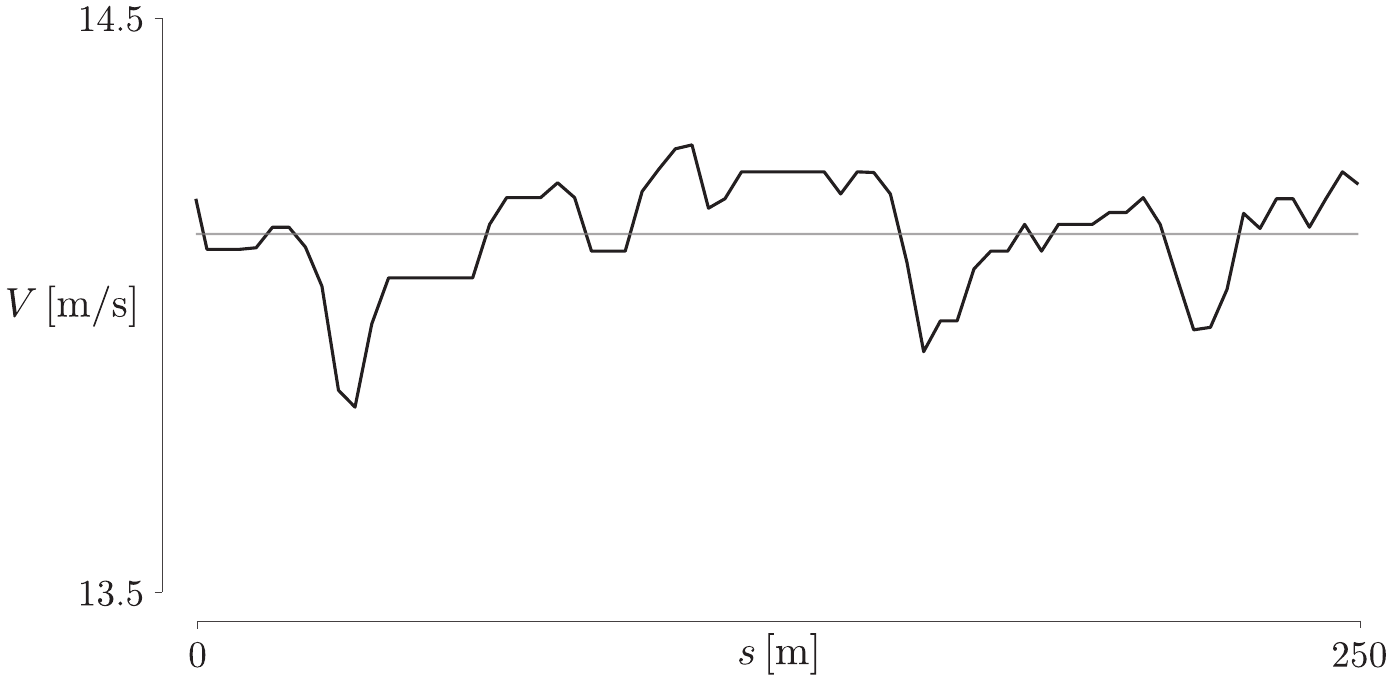}
\caption{\small Centre-of-mass speed and its average obtained from measurements}
\label{fig:FigLap6V}
\end{figure}
To gain insight into the empirical adequacy of the model, let us compare its predictions with measurements made by Mehdi Kordi ({\it pers.\!~comm.}, 2022).
Therein, we consider a steady lap, whose average power is $\overline P=402.8143\,\rm W$\,.
Other values are $m=80\,\rm kg$\,, $h=1\,\rm m$\,, $g=9.8126\,\rm m/s^2$\,, $\rho=1.1950\,\rm kg/m^3$\,.
The resistance coefficients are estimated to be ${\rm C_dA}=0.20\,{\rm m}^2$, ${\rm C_{rr}}=0.0025$\,, ${\rm C_{sr}}= 0.0035$ and $\lambda= 0.020$\,.

According to our model, $\overline P=\overline P_F+\overline P_U=375.3495\,{\rm W}+25.7459\,{\rm W}=401.0954\,{\rm W}$\,, which agrees with the measured power.
$\overline P_F$ alone would be an underestimate.

Also, measurements allow us to comment on the assumption of a constant centre-of-mass speed,~$V$.
Instantaneous $V$\,, obtained from measurements, is shown in Figure~\ref{fig:FigLap6V}\,.
Its mean is $\overline V=14.1211\,\rm m/s$ and its standard deviation is $0.0964\,\rm m/s$\,.
Only slight deviations from the mean and lack of a spatial pattern in instantaneous $V$, corresponding to segments of the track, support the aforementioned assumption.
These deviations might be mainly due to measurement errors.
\end{appendix}
\end{document}